# Declaration of Authorship

We hereby certify that this thesis is our own work and was completed for partial fulfilment of the undergraduate degree in Mechanical Engineering at Bangladesh University of Engineering and Technology. We declare that where any or all part of this thesis has previously been submitted for a degree or any other qualification at this university or any other university, this has been clearly stated. All the published work of others and all sources of information have been properly attributed and acknowledged and with the exception of such references, this work is entirely our own and all authors have contributed equally to the project.

Signed:

Md Imran Khan

________________________________

Md. Mamun Billah

________________________________

Mohammed Mizanur Rahman

________________________________

# Abstract

This thesis illustrates a numerical study of steady two dimensional mixed convention heat transfer phenomena in a rectangular channel with an active flow modification system by means of a variable speed heat conducting cylinder that is placed at the center of the channel. In the current study, an isothermal low temperature is maintained on the upper wall while a discrete isoflux heater is installed on the lower wall. The Prandtl number of air flowing through the channel is kept fixed at 0.71 while the Reynolds and the Grashof numbers are varied considerably for four different configurations of the cylinder such as: i) channel with no cylinder, ii) channel with stationary cylinder, iii) channel with cylinder rotating clockwise and iv) channel with counter clockwise rotating cylinder. The speed ratio is varied as 0.5, 1.0 and 1.5 for only the counter-clockwise and clockwise rotational setups. The impact on the flow pattern and temperature field is evaluated in terms of the distribution of streamlines, isothermal lines while the impact on the heat transfer phenomena is analyzed in terms of the local and surface averaged Nusselt numbers. The results point out that the cylinder's configuration, rotational speed, Grashof and Reynolds numbers significantly impact the flow pattern, temperature field and the heat transfer characteristics. Glapke and Afsaw [11] analyzed natural convection in a circular enclosure with a concentric hexagonal cylinder.

# Acknowledgements

All praises be to Allah, the Most Gracious and Most Merciful, for giving us the strength and knowledge to successfully complete this thesis.

We are extremely grateful and honored to have had the opportunity to work under the supervision of Dr. Mohammad Nasim Hasan, Associate Professor of Department of Mechanical Engineering, Bangladesh University of Engineering and Technology whose kindness, patience and guidance helped us shape our work into what it is today. For his mentorship and for the time and effort that was put in to make us understand the problem more clearly, we would like to express our sincerest gratitude and appreciation.

Dedicated to our beloved parents

# Table of Contents

# List of Figures

# List of Tables

# Nomenclature

| | |
|---|---|
| $g$ | gravitational acceleration |
| $C_p$ | constant pressure specific heat |
| $q$ | constant heat flux |
| $H$ | height of channel |
| $L$ | length of heat source |
| $Nu$ | Nusselt number |
| $p$ | dimensional pressure |
| $P$ | dimensionless pressure |
| $Pr$ | Prandtl number |
| $r$ | dimensional radius of the cylinder |
| $R$ | non-dimensional radius of the cylinder |
| $Re$ | Reynolds number |
| $Gr$ | Grashof number |
| $T$ | temperature |
| $x, y$ | Cartesian coordinates |
| $X,Y$ | non-dimensional Cartesian coordinates |
| $u, v$ | Cartesian velocity components |
| $U,V$ | non-dimensional Cartesian velocity components |
| $Nu_l$ | local Nusselt number |
| $Nu_{avg}$ | average Nusselt number |
| $Ra$ | Rayleigh number |

Greek symbols

| | |
|---|---|
| $\zeta$ | speed ratio |
| $\gamma$ | inclination angle |
| $\theta$ | dimensionless temperature |
| $\mu$ | dynamic viscosity |
| $\upsilon$ | kinematic viscosity |
| $\alpha$ | thermal diffusivity |
| $\rho$ | density |
| $\omega$ | angular rotating velocity |
| $\beta$ | thermal expansion coefficient |

Subscripts

| | |
|---|---|
| $o$ | value at the center of the cylinder |
| $c$ | cold |
| $s$ | solid |

# Chapter 1

## 1.1 Convection Heat Transfer

Convective heat transfer, often referred to simply as convection, is the transfer of heat from one place to another by the movement of fluids. Convection is usually the dominant form of heat transfer in liquids and gases. Although often discussed as a distinct method of heat transfer, convective heat transfer involves the combined processes of unknown conduction (heat diffusion) and advection (heat transfer by bulk fluid flow).

Convection can be "forced" by movement of a fluid by means other than buoyancy forces. Thermal expansion of fluids may also force convection. In other cases, natural buoyancy forces alone are entirely responsible for fluid motion when the fluid is heated, and this process is called "natural convection". An example is the draft in a chimney or around any fire. In natural convection, an increase in temperature produces a reduction in density, which in turn causes fluid motion due to pressures and forces when fluids of different densities are affected by gravity. For example, when water is heated on a stove, hot water from the bottom of the pan rises, displacing the colder denser liquid, which falls. After heating has stopped, mixing and conduction from this natural convection eventually result in a nearly homogeneous density, and even temperature. Without the presence of gravity, natural convection does not occur, and only forced-convection modes operate.

The convection heat transfer mode comprises one mechanism. In addition to energy transfer due to specific molecular motion (diffusion), energy is transferred by bulk, or macroscopic, motion of the fluid. This motion is associated with the fact that, at any instant, large numbers of molecules are moving collectively or as aggregates. Such motion, in the presence of a temperature gradient, contributes to heat transfer. Because the molecules in aggregate retain their random motion, the total heat transfer is then due to the superposition of energy transport by random motion of the molecules and by the bulk motion of the fluid. It is customary to use the term convection when referring to this cumulative transport and the term advection when referring to the transport due to bulk fluid motion.

Again, convective heat transfer can be classified as steady or unsteady depending on the time dependent properties of the fluid flow. The term steady implies that no change at a point with time. Unsteady flow is just the opposite of the steady flow with parameters being changed with time at any point in space. Also a flow is classified as being compressible and incompressible, depending on the level of variation of density during flow. A flow is said to be incompressible if the density remains nearly constant throughout. Gas flows can often be approximated as incompressible flow if the density changes are under about 5 percent, which is usually the case when Mach number, $M < 0.3$.

So, it can be concluded that convection heat transfer can be divided into three main categories on the basis of mechanism of sensible heat transfer as follows:

- Natural or free convection
- Forced convection
- Mixed convection or combined natural and forced convection

### 1.1.1 Natural Convection

Natural convection is a mechanism, or type of heat transport, in which the fluid motion is not generated by any external source (like a pump, fan, suction device, etc.) but only by density differences in the fluid occurring due to temperature gradients. In natural convection, fluid surrounding a heat source receives heat, becomes less dense and rises. The surrounding, cooler fluid then moves to replace it. This cooler fluid is then heated and the process continues, forming a convection current; this process transfers heat energy from the bottom of the convection cell to top. The driving force for natural convection is buoyancy, a result of differences in fluid density. Because of this, the presence of a proper acceleration such as arises from resistance to gravity, or an equivalent force (arising from acceleration, centrifugal force or Coriolis Effect), is essential for natural convection. For example, natural convection essentially does not operate in free-fall (inertial) environments, such as that of the orbiting International Space Station, where other heat transfer mechanisms are required to prevent electronic components from overheating.

Natural convection has attracted a great deal of attention from researchers because of its presence both in Nature and Engineering applications.

In nature, convection cells formed from air raising above sunlight-warmed land or water are a major feature of all weather systems. Convection is also seen in the rising plume of hot air from fire, oceanic currents, and sea-wind formation (where upward convection is also modified by Coriolis forces).

In engineering applications, convection is commonly visualized in the formation of microstructures during the cooling of molten metals, and fluid flows around shrouded heat-dissipation fins, and solar ponds. A very common industrial application of natural convection is free air cooling without the aid of fans: this can happen on small scales to large scale process equipment.

### 1.1.2 Forced Convection

Forced convection is a special type of heat transfer in which fluids are forced to move, in order to increase the heat transfer. This forcing can be done with a ceiling fan, a pump, suction device, or other.

Many people are familiar with the statement that "heat rises". This is a simplification of the idea that hot fluids are almost always less dense than the same fluid when cold, but there are exceptions (see the layers of the atmosphere and thermohaline circulation for exceptions). This difference in density makes hotter material naturally end up on top of cooler material due to the higher buoyancy of the hotter material.

Natural convection can create a noticeable difference in temperature within a home. Often this becomes places where certain parts of the house are warmer and certain parts are cooler. Forced convection creates a more uniform and therefore comfortable temperature throughout the entire home. This reduces cold spots in the house, reducing the need to crank the thermostat to a higher temperature, or putting on sweaters.

### 1.1.3 Mixed Convection

Mixed (combined) convection is a combination of forced and free convections which is the general case of convection when a flow is determined simultaneously by both an outer forcing system (i.e., outer energy supply to the fluid-streamlined body system) and inner volumetric (mass) forces, viz., by the non-uniform density distribution of a fluid medium in a gravity field. The most vivid manifestation of mixed convection is the motion of the temperature stratified mass of air and water areas of the Earth that the traditionally studied in geophysics. However, mixed convection is found in the systems of much smaller scales, i.e., in many engineering devices. We shall illustrate this on the basis of some examples referring to channel flows, the most typical and common cases. On heating or cooling of channel walls, and at the small velocities of a fluid flow that are characteristic of a laminar flow, mixed convection is almost always realized. Pure forced laminar convection may be obtained only in capillaries. Studies of turbulent channel flows with substantial gravity field effects have actively developed since the 1960s after their becoming important in engineering practice by virtue of the growth of heat loads and channel dimensions in modern technological applications (thermal and nuclear power engineering, pipeline transport).

## 1.2 Classification of Mixed Convection

Because of the wide range of variables, hundreds of papers have been published for experiments involving various types of fluids and geometries. This variety makes a comprehensive correlation difficult to obtain, and when it is, it is usually for very limited cases. Combined forced and natural convection, however, can be generally described in one of three ways.

### First case

The first case is when natural convection aids forced convection. This is seen when the buoyant motion is in the same direction as the forced motion, thus enhancing the heat transfer.

An example of this would be a fan blowing upward on a hot plate. Since heat naturally rises, the air being forced upward over the plate adds to the heat transfer.

### Second case

The second case is when natural convection acts in the opposite way of the forced convection. Consider a fan forcing air upward over a cold plate. In this case, the buoyancy force of the cold air naturally causes it to fall, but the air being forced upward opposes this natural motion, keeping the cool air hovering around the cold plate. This, in turn, diminishes the amount of heat transfer.

### Third case

The third case is referred to as transverse flow. This occurs when the buoyant motion acts perpendicular to the forced motion. This enhances fluid mixing, and enhances the heat transfer. An example of this is air flowing horizontally over a hot or cold pipe. This can encourage phase changes, which often creates a very high heat transfer coefficient. For example, steam leaving a boiler can pass through a pipe that has a fan blowing over it, cooling the steam back to a saturated liquid.

## 1.3 Mixed Convection in Cavity

Mixed convection can also be classified as internal flow and external flow. When a fluid is surrounded by a solid boundary, the flow can be termed as the internal flow. An example of internal flow is liquid flowing through a pipe. On the contrary, external flow is observed when a fluid is open to its surroundings and extends indefinitely without colliding a solid boundary. Now a days many investigations have been made on mixed convection in cavity both experimentally and numerically. In many cases, excellent and satisfactory agreement is obtained between the two results.

### 1.3.1 Closed Cavity Problems

When a system is comprised in such a way that it is closed and no mass transfer occurs from or to the cavity, then these problems are termed as closed cavity problems. A conveyor belt removing heat from an enclosed area can be an effective practical application of closed cavity problems.

### 1.3.2 Open Cavity Problems

When the system remains open and the mass transfer occurs from or to the cavity, then these problems are termed as open cavity problems. So, these problems are just the opposite of closed cavity problems. In these problems, source of forced flow is external flow from inlet to outlet of an enclosure.

## 1.4 Literature Review

Steady state mixed convection phenomena in two dimensional is of prime concern in many applications that ranges from the development of advanced cooling devices, heat exchangers, micro electronic systems, etc. In many cases, rotating obstacles are used to enhance the rate of cooling in such devices and nowadays this has gained popularity. Chang *et al.* [1] conducted numerical simulation in an irregular two-dimensional cavity generated from an outer circular envelope and a square cylinder on the inside. Salam *et al.* [4] utilized a square enclosure with a rotating circular cylinder placed at different vertical positions to study the mixed convection phenomena. Fu *et al.*[6] demonstrated that the cylinder's direction of rotation has a significant impact on the heat transfer characteristics. Costa *et al.*[8] used insulated vertical walls in a heated square enclosure with a rotating cylinder placed at different heights. Oosthuizen and Paul [9] investigated this phenomena in a cavity with a heated half cylinder kept at a high isothermal temperature. Glapke and Afsaw [11] showed findings for natural convection in a circular enclosure equipped with a concentric hexagonal cylinder.Ha *et al.* [12] and [20] utilized a square cavity with adiabatic side walls and isothermal upper and bottom walls. Fu and Tong [14] analysed heat transfer and flow patterns with a heated, oscillating cylinder. Guimarães and Menon [15] employed several discrete heaters to analyze mixed convection in a rectangular channel. Müftüoğlu and E. Bilgen [16] showed that when the heaters are placed at the bottom, a high degree of heat transfer performance is achieved. House *et al.* [17] used a vertical, square enclosure with a centered heat conducting body as an obstruction to the flow of fluid. Ghadder and Thiele [18] and Ghadder [21] utilizedthe cylinder as a heat source in the analysis of their rectangular cavity. Elepano and Oosthuizen [19] analyzed mixed convection in a cavity wih a cooled top surface and a heated cylinder. Asan [22] studied two-dimensional natural convection and concluded that the dimension ratio and Rayleigh have a noteworthy effect on the distribution of isothermal lines and streamlines. Yoo and Joo-sik [23] investigated mixed convection with air between two horizontal concentric cylinders held at different uniform temperature.The cold outer cylinder causes forced flow and rotates slowly with at a fixed velocity. Roslan [24] analyzed natural convection in a cavity with a sinusoidal heated cylindrical source. He concluded the oscillation of the cylinder temperature had a major impact on the isotherm and streamline pattern. Aminossadati and Ghasemi [25] investigated the cooling caused by natural convection in a square cavity occupied by a Nano fluid and equipped with a heat source at the bottom boundary. The results indicate that the rise of Rayleigh number enhances heat transfer with a fall in temperature of the heat source. Increasing the length of the heat source augments the transfer of heat to the working fluid and cause the surface temperature of the heat source to rise. Shuja *et al.*[26] carried out a numerical investigation in a square enclosure with heat generating rectangular obstacle and took into effect the impact of the location of outlet ports. Thermo-fluid analysis was carried out by the authors Khan et al. [27] and Billah et al. [28] for a thermal system, and this thesis deeply investigates and illustrates the findings of the earlier investigation into understanding the cooling effect inside a rectangular channel.

# Chapter 2

## Mathematical Modeling

### 2.1 Physical Model

The schematic diagram of the setup is shown in figure one. A discrete isoflux heater with constant heat flux, *q* and length, *L* is placed on the bottom wall. Here *L* = *H*. The top wall, inlet and outlet are kept at a low isothermal temperature, $T_c$. A rotating heat conducting cylinder ($R$=0.15$H$), which rotates at 3 speed ratios is placed at the center of the channel. The region of the bottom wall without the heater is considered adiabatic. The no slip condition is applicable on the surface of the cylinder and the walls of the channel. The Grashof number is varied between $10^4$ and $10^5$ while the Reynolds number is changed between 10 and 100. The Prandtl number of air is set to 0.71. The solid-fluid conductivity ratio, K = 10 with a zero pressure boundary condition at the outlet.

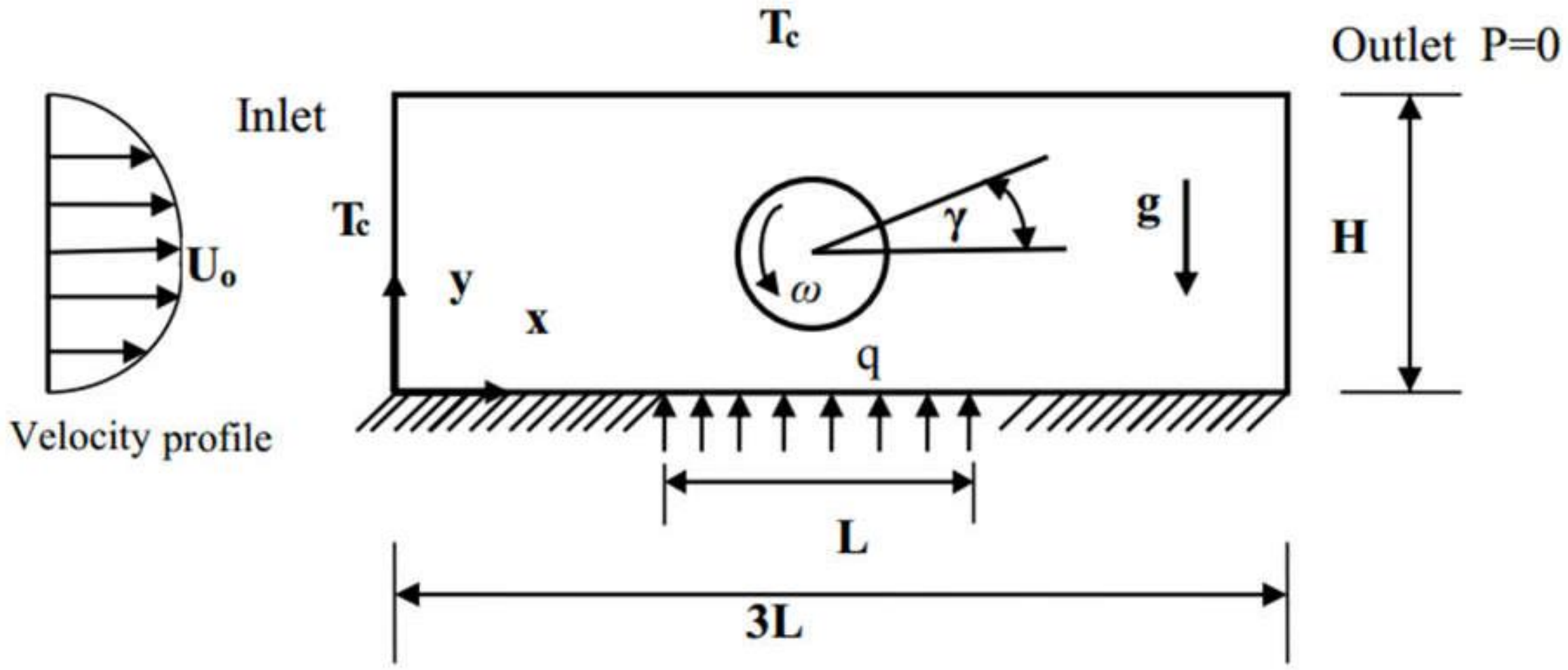

Figure 1: Schematic of the configuration

## 2.2 Assumptions

a. Heat transfer by radiation is negligible
b. Steady state analysis is applicable which means that there is no change of velocity with time
c. Boussinesq approximation is used for the density in the buoyancy term
d. No slip condition is applicable on the cylinder surface and channel walls
e. The fluid is incompressible
f. Heat and fluid flow is two dimensional
g. The fluid properties are not direction dependent i.e. the fluid is isotropic and fluid flow is laminar

## 2.3 Governing Equations

The two-dimensional steady state continuity, momentum and energy equation are used in modelling the present problem for the flow and thermal fields. Based on the above assumptions, the dimensional governing equations are:

$$\frac{\partial u}{\partial x}+\frac{\partial v}{\partial y}=0 \tag{2.1}$$

$$u\frac{\partial u}{\partial x}+v\frac{\partial u}{\partial y}=-\frac{1}{\rho}\frac{\partial p}{\partial x}+\nu\left(\frac{\partial^2 u}{\partial x^2}+\frac{\partial^2 u}{\partial y^2}\right) \tag{2.2}$$

$$u\frac{\partial v}{\partial x}+v\frac{\partial v}{\partial y}=-\frac{1}{\rho}\frac{\partial p}{\partial y}+\nu\left(\frac{\partial^2 v}{\partial x^2}+\frac{\partial^2 v}{\partial y^2}\right)+g\beta\left(T-T_c\right) \tag{2.3}$$

$$u\frac{\partial T}{\partial x}+v\frac{\partial T}{\partial y}=\alpha\left(\frac{\partial^2 T}{\partial x^2}+\frac{\partial^2 T}{\partial y^2}\right) \tag{2.4}$$

The energy equation for the heat conducting cylinder is,

$$\frac{\partial^2 T_s}{\partial x^2}+\frac{\partial^2 T_s}{\partial y^2}=0 \tag{2.5}$$

## 2.4 Dimensional Analysis

The dimensional governing equations and parameters must be made non-dimensional. Thus the dimensionless governing equations are:

$$\frac{\partial U}{\partial X}+\frac{\partial V}{\partial Y}=0 \tag{2.6}$$

$$U\frac{\partial U}{\partial X}+V\frac{\partial U}{\partial Y}=-\frac{\partial P}{\partial X}+\frac{1}{\mathrm{Re}}\left(\frac{\partial^2 U}{\partial X^2}+\frac{\partial^2 U}{\partial Y^2}\right) \tag{2.7}$$

$$U\frac{\partial V}{\partial X}+V\frac{\partial V}{\partial Y}=-\frac{\partial P}{\partial Y}+\frac{1}{\mathrm{Re}}\left(\frac{\partial^2 V}{\partial X^2}+\frac{\partial^2 V}{\partial Y^2}\right)+\frac{Gr}{\mathrm{Re}^2}\theta \tag{2.8}$$

$$U\frac{\partial \theta}{\partial X}+V\frac{\partial \theta}{\partial Y}=\left(\frac{\partial^2 \theta}{\partial X^2}+\frac{\partial^2 \theta}{\partial Y^2}\right) \tag{2.9}$$

The energy equation for the heat conducting cylinder is:

$$\frac{\partial^2 \theta_s}{\partial X^2}+\frac{\partial^2 \theta_s}{\partial Y^2}=0 \tag{2.10}$$

The non-dimensional parameters are:

$$X=\frac{x}{L},\ Y=\frac{y}{L},\ U=\frac{u}{U_{\max}},\ V=\frac{v}{U_{\max}},\ \theta=\frac{T-T_c}{qL/k},\ \mathrm{Pr}=\frac{\nu}{\alpha},\ Gr=\frac{g\beta qL^4}{\nu^2 k},\ P=\frac{p}{\rho_0 U_0^2},\ \mathrm{Re}=\frac{U_{\max}L^2}{\nu}$$

### 2.4.1 Dimensionless Numbers

A number without a physical unit attached to it is defined as a dimensionless number or parameter. These are formed from dimensional number but ultimately the units cancel out. The dimensionless numbers used in this study are given below:

Reynolds Number: It is defined as the ratio of inertia to viscous forces. It determines whether the flow is laminar or turbulent.

Prandtl Number: It is defined as the ratio of molecular diffusivity of momentum to molecular diffusivity of heat. It can also be used as the ratio of the hydrodynamic and thermal boundary layer.

Nusselt Number: It is the ratio of convective to conductive heat transfer across (normal to) the boundary. It represents the enhancement of heat transfer through a fluid layer as a result of convection relative to conduction.

Grashof Number: It is defined as the ratio of buoyancy force to viscous force acting on the fluid.

Rayleigh Number: It is defined as the product of Grashof number and Prandtl number.

Richardson Number: It represents the importance of natural convection relative to forced convection.

## 2.5 Dimensionless Boundary Conditions

The dimensionless boundary condition are given below:

At the top wall: $U = 0, V = 0, \theta = 0$

At the inlet: $\theta = 0$ , $\frac{U_0}{U_{\max}} = 4Y(1 - Y)$

At the outlet: $p = 0$

At the bottom wall where it is adiabatic: $\frac{\partial \theta}{\partial Y} = 0$

At the heated section of the bottom wall: $\frac{\partial \theta}{\partial Y} = -1$

At the cylinder surface: $U = -\frac{U_p}{U_{\max}} \sin\gamma$, $V = \frac{U_p}{U_{\max}} \cos\gamma$, $0 \le \gamma \le 2\pi$ where $U_p$ = peripheral speed.

$$U = -\zeta \sin\gamma,\ V = \zeta \cos\gamma \tag{2.11}$$

Here, $U$ is negative for the clockwise case. For the anti-clockwise case, $V$ is negative.

At the solid-fluid interface of the cylinder: $U = 0, V = 0$ and $\left(\frac{\partial \theta}{\partial N}\right)_{fluid} = K\left(\frac{\partial \theta}{\partial N}\right)_{solid}$

Where $N$ is the non-dimensional distance and is either $X$ or $Y$ and $K = K_s / K_f$ is the solid-fluid conductivity ratio.

Local Nusselt number is:

$$Nu_l = \frac{1}{\theta(X)} \tag{2.12}$$

The average Nusselt number is:

$$Nu_{avg} = \frac{1}{\varepsilon}\int_0^{\varepsilon} \frac{1}{\theta(X)} dX \tag{2.13}$$

Where $1 \le X \le 2$ and $0 \le \varepsilon \le 1$

The flow field is evaluated by the distribution of the streamlines that are given by the stream function $\psi$ which is obtained from the velocity components $U$ and $V$. The stream function is defined in such a way that:

$$U = \frac{\partial \psi}{\partial Y}, V = -\frac{\partial \psi}{\partial X} \tag{2.14}$$

$$\frac{\partial^2 \psi}{\partial X^2} + \frac{\partial^2 \psi}{\partial Y^2} = \frac{\partial U}{\partial Y} - \frac{\partial V}{\partial X} \tag{2.15}$$

The boundary conditions that are applied to solve the governing equation includes the no slip condition that exists at the wall and on the surface of the cylinder. As per convention $\psi$ is considered positive for circulation in counter-clockwise rotation and vice-versa.

# Chapter 3

## Numerical Methodology

Numerical analysis is the study of algorithms that use numerical approximation (as opposed to general symbolic manipulations) for the problems of mathematical analysis (as distinguished from discrete mathematics).Numerical analysis naturally finds applications in all fields of engineering and the physical sciences, but in the 21st century also the life sciences and even the arts have adopted elements of scientific computations. Ordinary differential equations appear in celestial mechanics (planets, stars and galaxies); numerical linear algebra is important for data analysis; stochastic differential equations and Markov chains are essential in simulating living cells for medicine and biology.

Before the advent of modern computers numerical methods often depended on hand interpolation in large printed tables. Since the mid-20th century, computers calculate the required functions instead. These same interpolation formulas nevertheless continue to be used as part of the software algorithms for solving differential equations.

To obtain an approximate solution numerically, we have to approximate the partial differential equations by a system of algebraic equations, which can be solved on a computer. Actually numerical analysis is the study of algorithms that uses some approximations to simplify the analysis. The approximations are applied to small domains in space and the numerical solution provides results at discrete locations in space.

## 3.1 Advantages of Numerical Analysis

Many researchers have performed experimental investigation on the fluid flow, heat and mass transfer phenomena. But, the experimental approach is quite costly in many cases and time consuming which may not be desirable. To reduce the problem, numerical analysis becomes a very attractive approach today. On the other hand, this technique provides accurate results with reduced time effort. The key advantages of the numerical analysis are mentioned below:

1. When analytical solution of the mathematically defined problem is possible but it is time-consuming and the error of approximation we obtain with numerical solution is acceptable. In this case the calculations are mostly made with use of computer because otherwise it is highly doubtful if any time is saved. It is also done individually to decide what we mean by "time-consuming analytical solution". In my discipline even very simple mechanical problems are solved numerically simply because of laziness.

2. When analytical solution is impossible which means that we have to apply numerical methods in order to find the solution. This does not mean that we must do calculations with computers although it usually happens so because of the number of required operations.

3. This technique converts complex geometry into simpler ones by discretizing the physical problem into small domains.

4. Lower cost is involved in the numerical procedure than experimental approach.

## 3.2 Solution of Differential Equations

The mathematical model of complex geometry may be of very large shapes and limitations arise in this case for numerical modeling. So, scientists and researchers have sub-divided the complex system into smaller components or elements and thus the complex model can be analyzed easily. Based on the way of discretization different types of solution methods are available such as,

- Finite difference method (FDM)
- Finite element method (FEM)
- Finite volume method (FVM)

There are numerous other methods such as SEM, LBM etc. to solve a problem in CFD. And, many software packages such as ANSYS, COMSOL Multiphysics, and MARC etc. are available now a days. In this work, we used COMSOL Multiphysics for simulation.

## 3.3 Finite Element Method

The description of the laws of physics for space- and time-dependent problems are usually expressed in terms of partial differential equations (PDEs). For the vast majority of geometries and problems, these PDEs cannot be solved with analytical methods. Instead, an approximation of the equations can be constructed, typically based upon different types of discretization. These discretization methods approximate the PDEs with numerical model equations, which can be solved using numerical methods. The solution to the numerical model equations are, in turn, an approximation of the real solution to the PDEs. The finite element method (FEM) is used to compute such approximations.

For instance, a function $u$ that may be the dependent variable in a PDE (i.e., temperature, electric potential, pressure, etc.). The function $u$ can be approximated by a function $u_h$ using linear combinations of basis functions according to the following expressions:

$U=U_h$

And

$U_h=\sum U_i\psi_i$

Here, $\psi_i$ denotes the basis functions and $u_i$ denotes the coefficients of the functions that approximate $u$ with $u_h$. The figure below illustrates this principle for a 1D problem. $u$ could, for instance, represent the temperature along the length (x) of a rod that is non-uniformly heated. Here, the linear basis functions have a value of 1 at their respective nodes and 0 at other nodes. In this case, there are seven elements along the portion of the x-axis, where the function u is defined (i.e., the length of the rod).

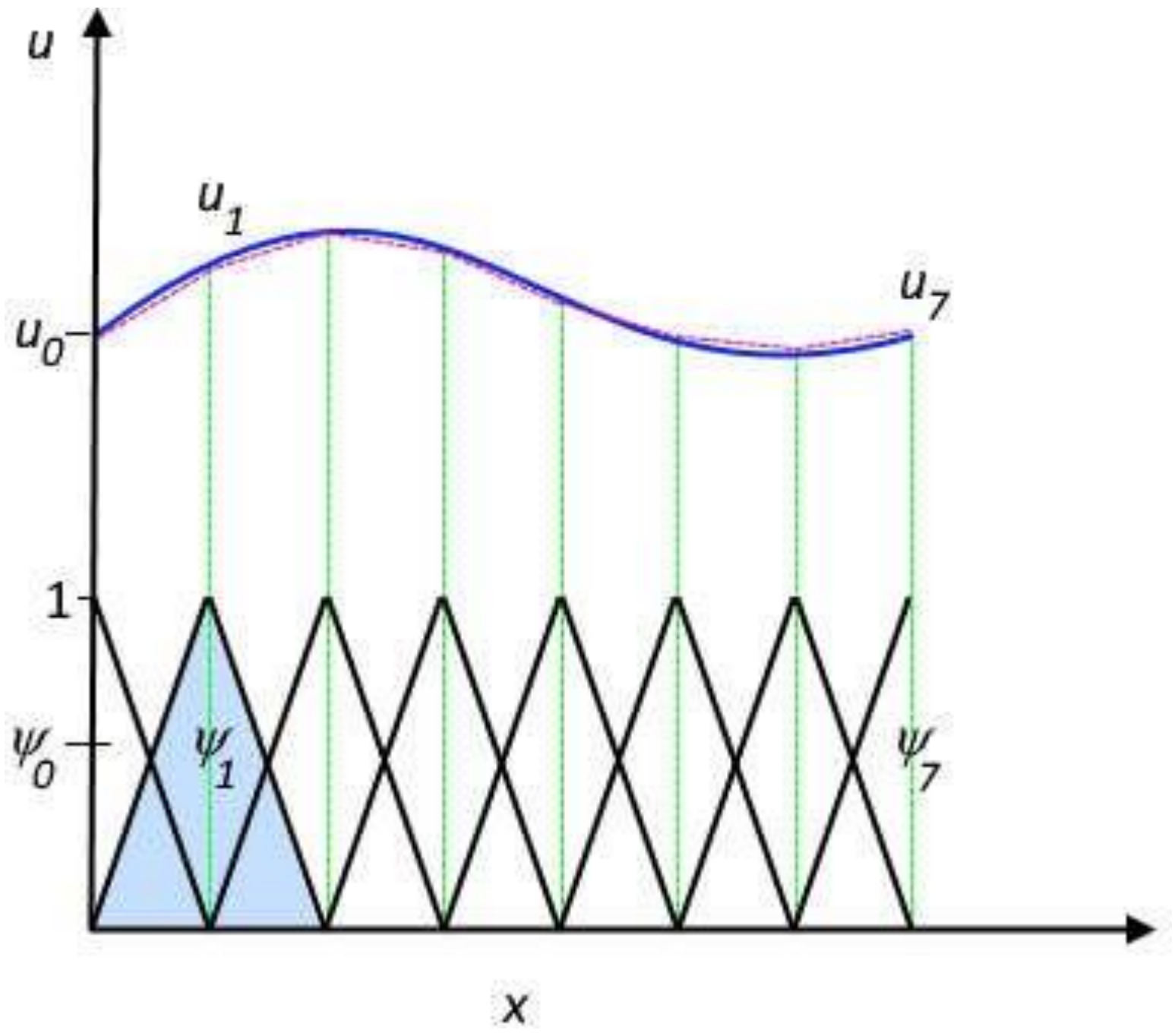

Figure 2: Finite element discretization of a domain by dividing it into 7 elements

The function *u* (solid blue line) is approximated with $u_h$ (dashed red line), which is a linear combination of linear basis functions ($\psi_i$ is represented by the solid black lines). The coefficients are denoted by $u_0$ through $u_7$.

One of the benefits of using the finite element method is that it offers great freedom in the selection of discretization, both in the elements that may be used to discretize space and the basis functions. In the figure above, for example, the elements are uniformly distributed over the x-axis, although this does not have to be the case. Smaller elements in a region where the gradient of *u* is large could also have been applied, as highlighted below.

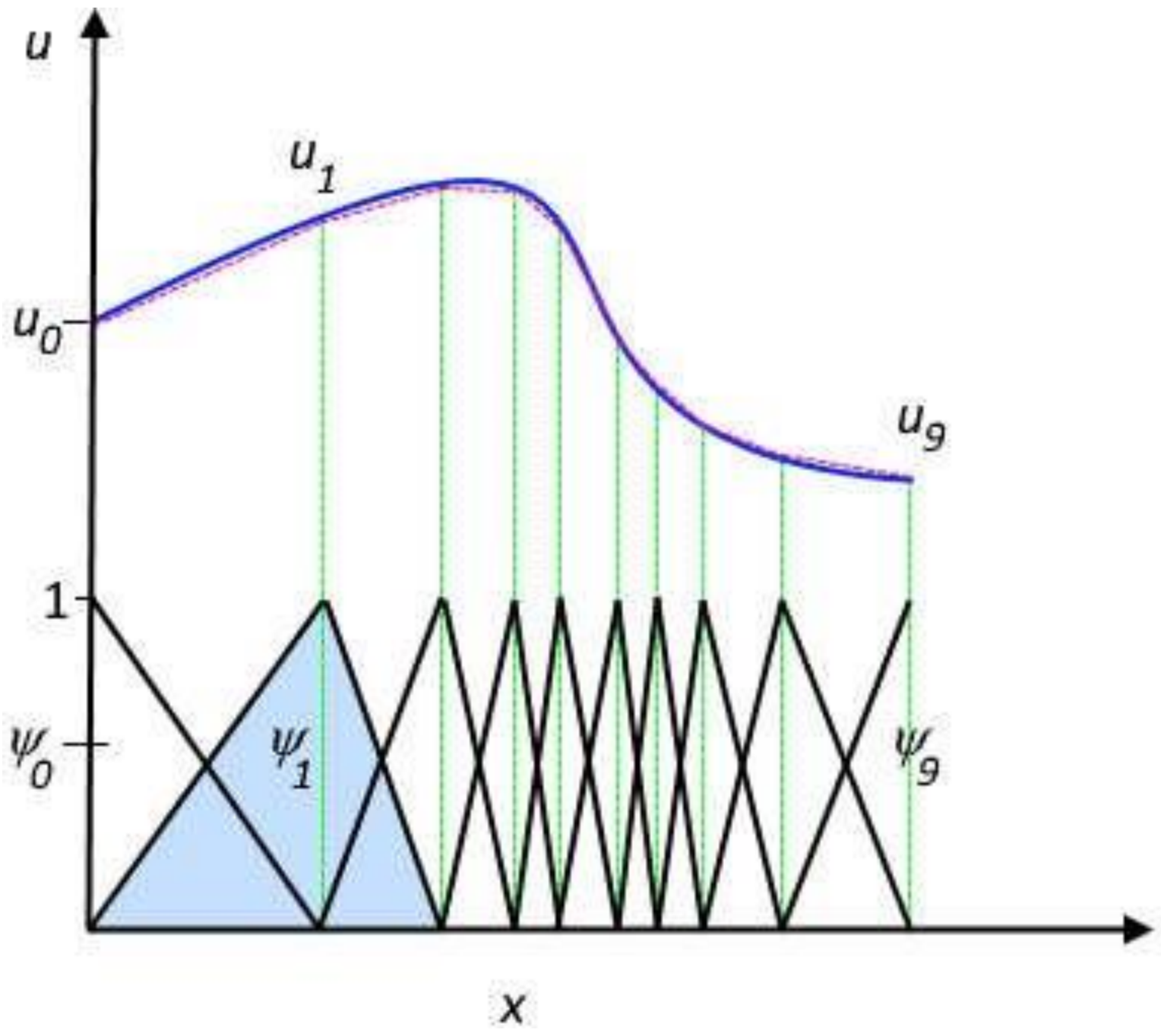

Figure 3: Finite element discretization of a domain by dividing it into 9 elements

Both of these figures show that the selected linear basis functions include very limited support (nonzero only over a narrow interval) and overlap along the x-axis. Depending on the problem at hand, other functions may be chosen instead of linear functions.

Another benefit of the finite element method is that the theory is well developed. The reason for this is the close relationship between the numerical formulation and the weak formulation of the PDE problem. For instance, the theory provides useful error estimates, or bounds for the error, when the numerical model equations are solved on a computer.

Looking back at the history of FEM, the usefulness of the method was first recognized at the start of the 1940s by Richard Courant, a German-American mathematician. While Courant recognized its application to a range of problems, it took several decades before the approach was applied generally in fields outside of structural mechanics, becoming what it is today.

### 3.3.1 Advantages of Finite Element Method

- Comprehensive result sets, generating the physical response of the system at any location, including some which might have been neglected in an analytical approach.
- Safe simulation of potentially dangerous, destructive or impractical load conditions and failure modes.
- Optimal use of a model. Often, several failure modes or physical events can be tested within a common model.
- The simultaneous calculation and visual representation of a wide variety of physical parameters such as stress or temperature, enabling the designer to rapidly analyses performance and possible modifications.
- Extrapolation of existing experimental results via parametric analyses of validated models.
- Relatively low investment and rapid calculation time for most applications.

## Some disadvantages may also be included

- For large number of elements the standard analysis from finite element methods do not yield useful results, often leading to "pollution" errors. There are some ways to improve this, but this is still an active area of research. More generally this means that a nonlinear reaction term that leads to a large Jacobian will probably run into misleading results.
- A general closed-form solution (An equation is said to be a closed-form solution if it solves a given problem in terms of functions and mathematical operations from a given generally-accepted set. For example, an infinite sum would generally not be considered closed-form.) which would permit one to examine system response to changes in various parameters, is not produced.
- The FEM obtains only "approximate" solutions.
- Mistakes by users can be fatal.

## 3.4 Galerkin Weighted Residual Method

In finite element method, there are various methods to obtain the approximate solution to the given problem. The methods are:

- Ritz method
- Rayleigh Ritz method
- Collocation method
- Weighted Residual method

In the present work, problems are investigated using **Galerkins weighted residual method** which is the mostly used and most powerful technique used in FEM. So, main emphasis will be given on this method.

This method is chosen because the global system matrix is decomposed into smaller matrices and then these sub-matrices are solved using a non-linear parametric solver. Six node non-uniform triangular mesh elements are used in this work as it results smooth non-linear variations of field variables and the used method ensures fast convergence and also the reliability.

Galerkin weighted residual method is an approximate analytical method suitable for direct solution to differential equations. This technique involves the principle of weighted residuals. It has wider scope of using as a tool to FEM formulation compared to Rayleigh-Ritz method. This method is very effective for practical applications as solutions can be directly obtained from the governing differential equation and boundary conditions and this method does not need to derive the functional compared to Rayleigh-Ritz method. Method of weighted residual actually takes cue from variational principle to solve problem for the entire domain. The philosophy behind this method is very straightforward. It is known that, for numerically solving a problem, an approximate solution is at first assumed. Now, if one puts the approximated value of solution into the governing equation, it will not satisfy the equation completely like an exact solution. There will always be a residue which is the difference between the values of dependent variable produced for exact solution and approximate solution. Mathematically,

$$y_{exact} - y_{approximate} = R$$

Here, $R$ is the residual. As numerically calculated solution comes closer to the exact solution, value of $R$ tends to zero. So the ultimate target of solution should be to minimize $R$ essentially to zero. General weighted residual formulation is,

$$\int_{\Omega} w_i(x) R dx = 0$$

Where $i$= 1, 2, 3, ……,n Functions $w_i(x)$ are called the Residual function or Error function. Purpose of this weighting function is to modify the values of residuals of each element in such way that when integrated over all the elements of a domain (Here, dx is element for one dimensional domain and Ω represents integral over entire domain), value of the integral is zero. This particular function is the unique feature of this method and it gets its name from this function. To solve a problem with weighted residual method, one is free to choose a weighting function. This can be a Dirac-Delta function or a simple polynomial function. However, some particular forms of the weighting function are found to be convenient in numerical calculation.

In Galerkin Weighted residual method, the weighting function is assumed to the trial function which is used to get the approximate solution of the equations over each element. Here “similar” does not mean that the function must be exactly similar, rather, these should be of the same order and form. Therefore, for Galerkin weighted residual method the equation is,

$$\int_{\Omega} N_i(x) R dx = 0$$

Where, $N_i(x)$ is the weighting function. Important properties of the weighting functions are mentioned below,

- Weighting function should satisfy homogeneous parts of essential boundary conditions of governing differential equations.
- Weighting function must be continuous.

- Value of shape functions $N_i$ is 1 at node $i$, whereas, this value is dependent on the coordinates ($x$, $y$) for two dimensional element.
- $\sum_{i}^{6} N_i = 1$ For six-node triangular element.

## 3.5 Mesh Generation

Mesh generation is the practice of generating a polygonal or polyhedral mesh that approximates a geometric domain. The term "grid generation" is often used interchangeably. Typical uses are for rendering to a computer screen or for physical simulation such as finite element analysis or computational fluid dynamics. The input model form can vary greatly but common sources are CAD, NURBS, B-rep, STL or a point cloud. The field is highly interdisciplinary, with contributions found in mathematics, computer science, and engineering.

Three-dimensional meshes created for finite element analysis need to consist of tetrahedral, pyramids, prisms or hexahedra. Those used for the finite volume method can consist of arbitrary polyhedral. Those used for finite difference methods usually need to consist of piecewise structured arrays of hexahedra known as multi-block structured meshes. A mesh is otherwise a discretization of a domain existing in one, two or three dimensions.

The entire domain of square cavity is discretized into mesh elements of different sizes. Discretization is possible for variety of element shapes such as triangles, quadrilaterals etc. There are certain number of nodes at different points of the elements. The choice of interpolation function and the number of nodes determine the solution pattern. Suitable known functions, representing approximate solution, are assumed over each element. Order of the solution is dependent on the nodes of element.

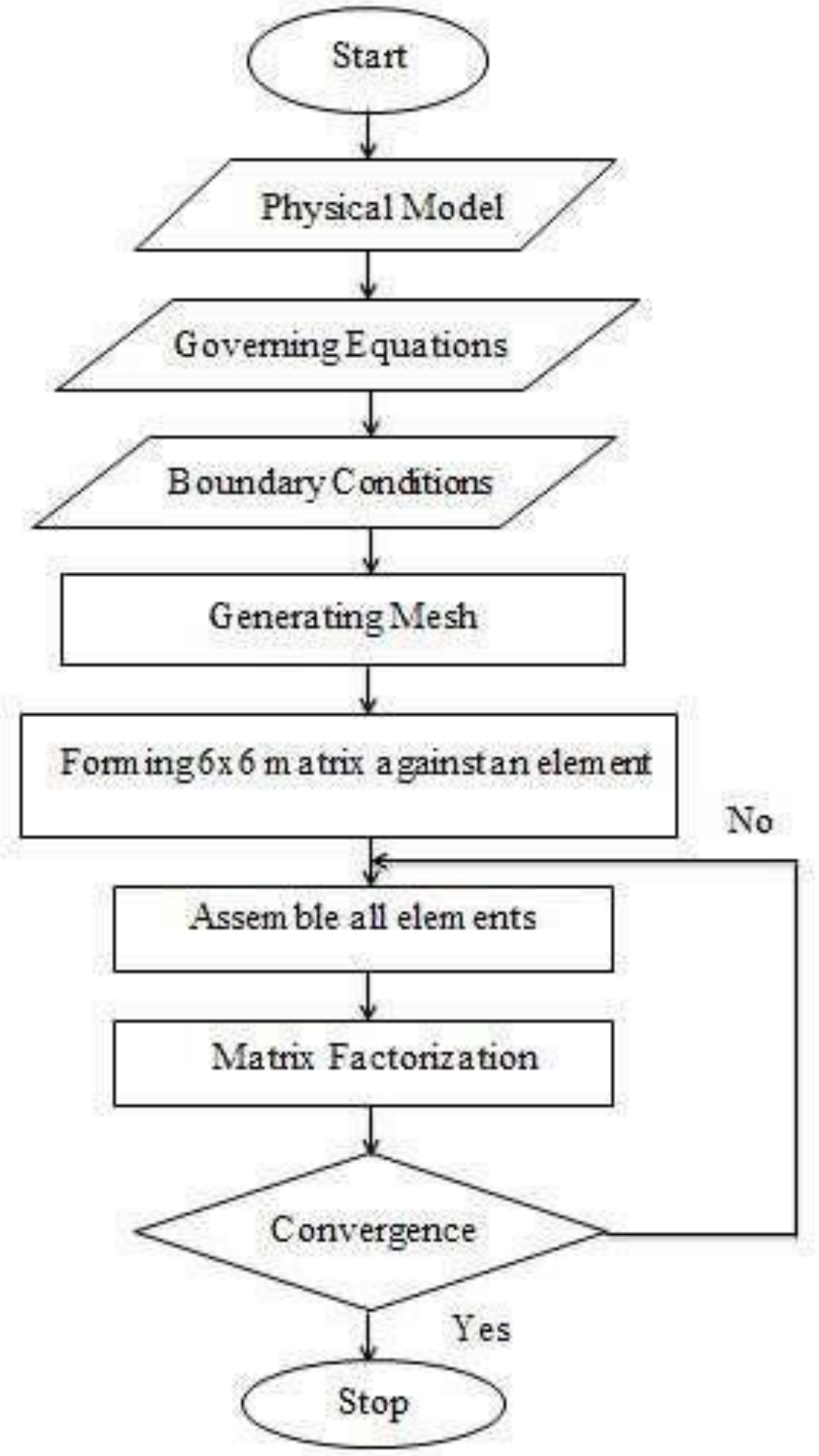

Figure 4: Flowchart of Finite Element Method

Most of the fluid flow equations are easily solved by discretizing procedures using the Cartesian coordinate system. In this system the implementation of finite volume method is simpler and easier to understand. But most of the engineering problems deal with complex geometries that do not work well in the Cartesian coordinate system. When the boundary region of the flow doesn't coincide with the coordinate lines of the structured grid then we can solve the problem by geometry approximation. The following figure shows how a cylinder can be approximated with the Cartesian coordinate system.

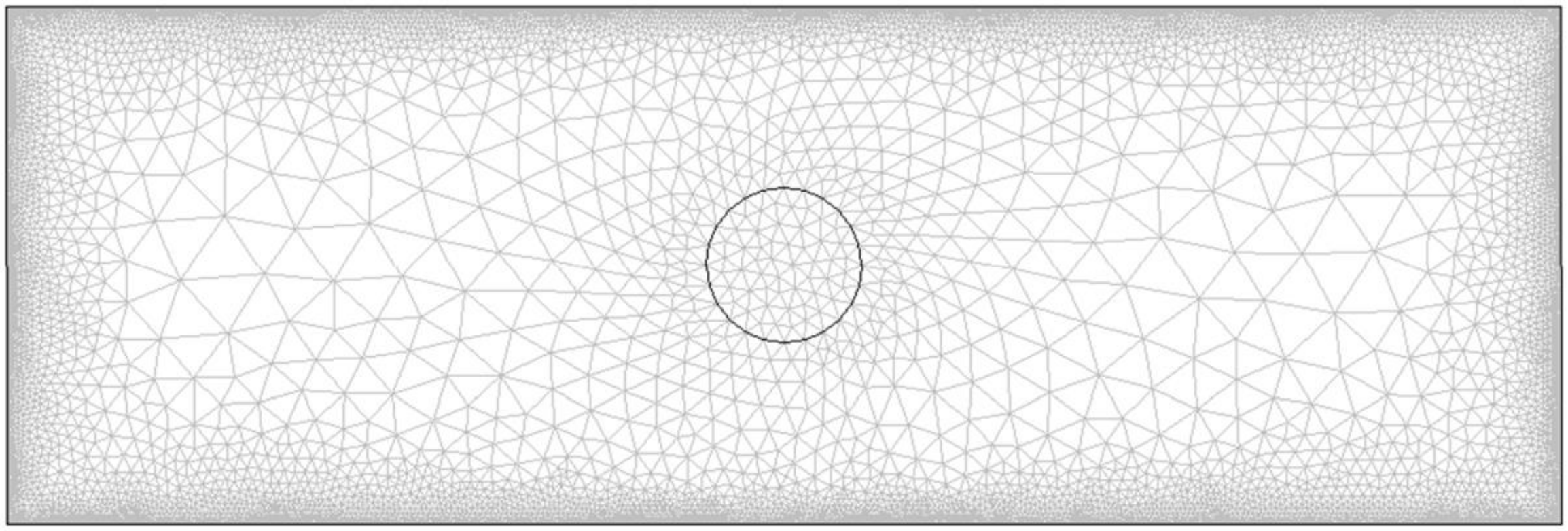

Fig 5: Finite element mesh generation inside the computational domain of rectangular enclosure.

## 3.6 Grid Sensitivity Test

In order to obtain grid independent solution, a grid refinement study is performed for a square enclosure with the horizontal conductive circular cylinder at Re=100, Gr=10 and K=10. In the present work, six combinations of non-uniform grids are used to test the effect of grid size on the accuracy of the predicted results. The following figure shows the convergence of the average Nusselt number ($Nu_{avg}$), at the heated bottom wall with grid refinement. It is observed that grid independence is achieved with grid size of 9878 control volume where there is insignificant change in the average Nusselt number ($Nu_{avg}$) with the improvement of finer grid. The agreement is found to be excellent which validates the present computations indirectly. Hence, for the rest of the calculation in this study, a grid size of 9878 control volume is chosen for optimum results.

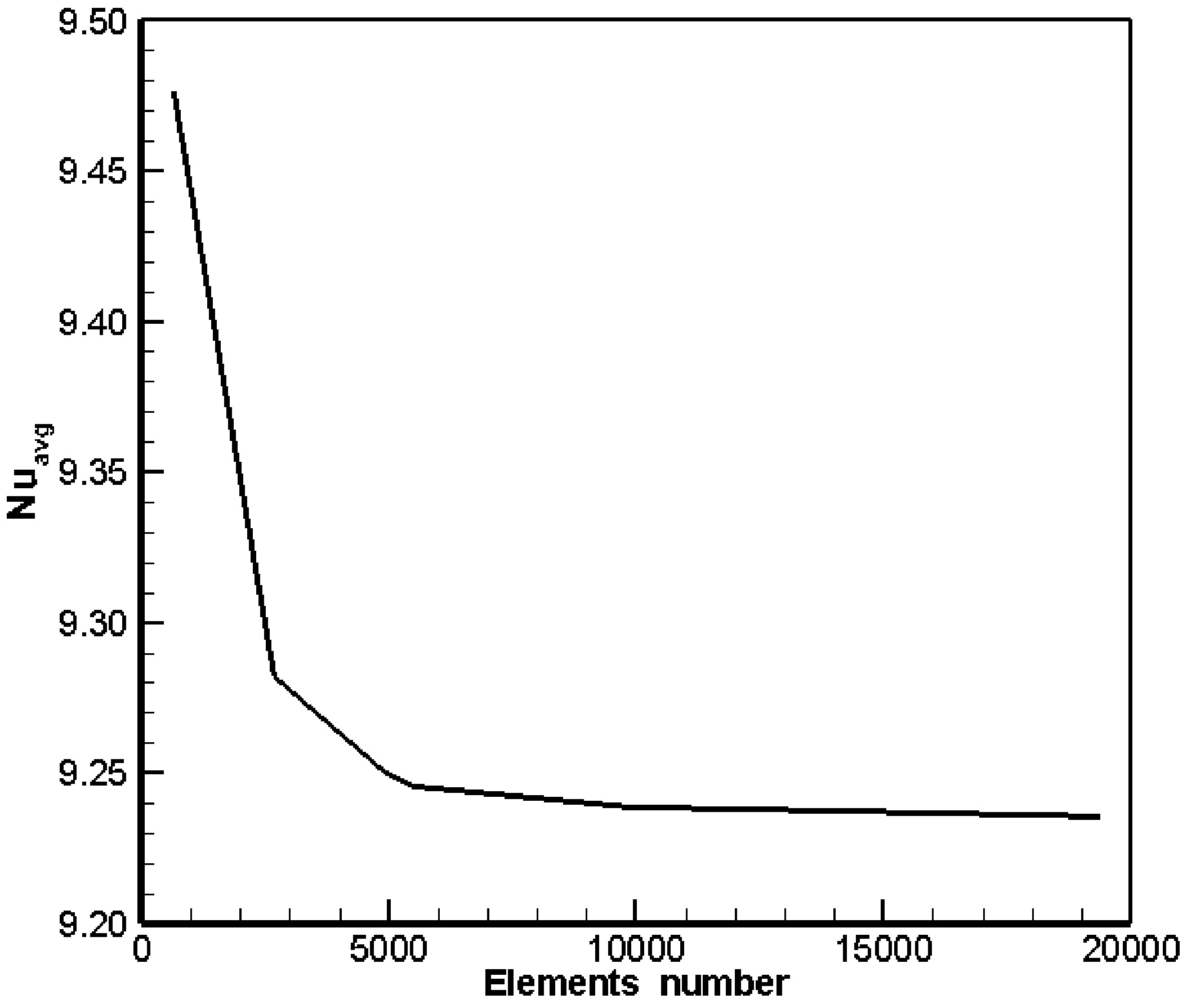

Fig. 6: Variation of average Nusselt number with different mesh size.

## 3.7 Validation of Isothermal Lines and Streamlines

To check the adequacy of the numerical scheme, the natural convection problem for a low temperature outer square enclosure and high temperature inner circular cylinder was tested. The calculated surface-averaged Nusselt numbers for the test case are compared with the benchmark values by Kim *et al.* [2] as shown in Table 1. Further verification was performed by using the present numerical algorithm to investigate the same problem considered by Kim *et al.* [2] using the same flow conditions and geometries, which were reported for laminar natural convection heat transfer using the same boundary conditions but the numerical scheme is different. The comparison is made using the following dimensionless parameters: $Pr$ = 0.71, $Ra$ =$10^6$. Excellent agreement was achieved between Kim *et al.* [30] results and the present numerical scheme results for both the streamlines and temperature contours inside the square enclosure with the inner cylinder along the horizontal centerline as shown in the following figure. These verifications make a good confidence in the present numerical model.

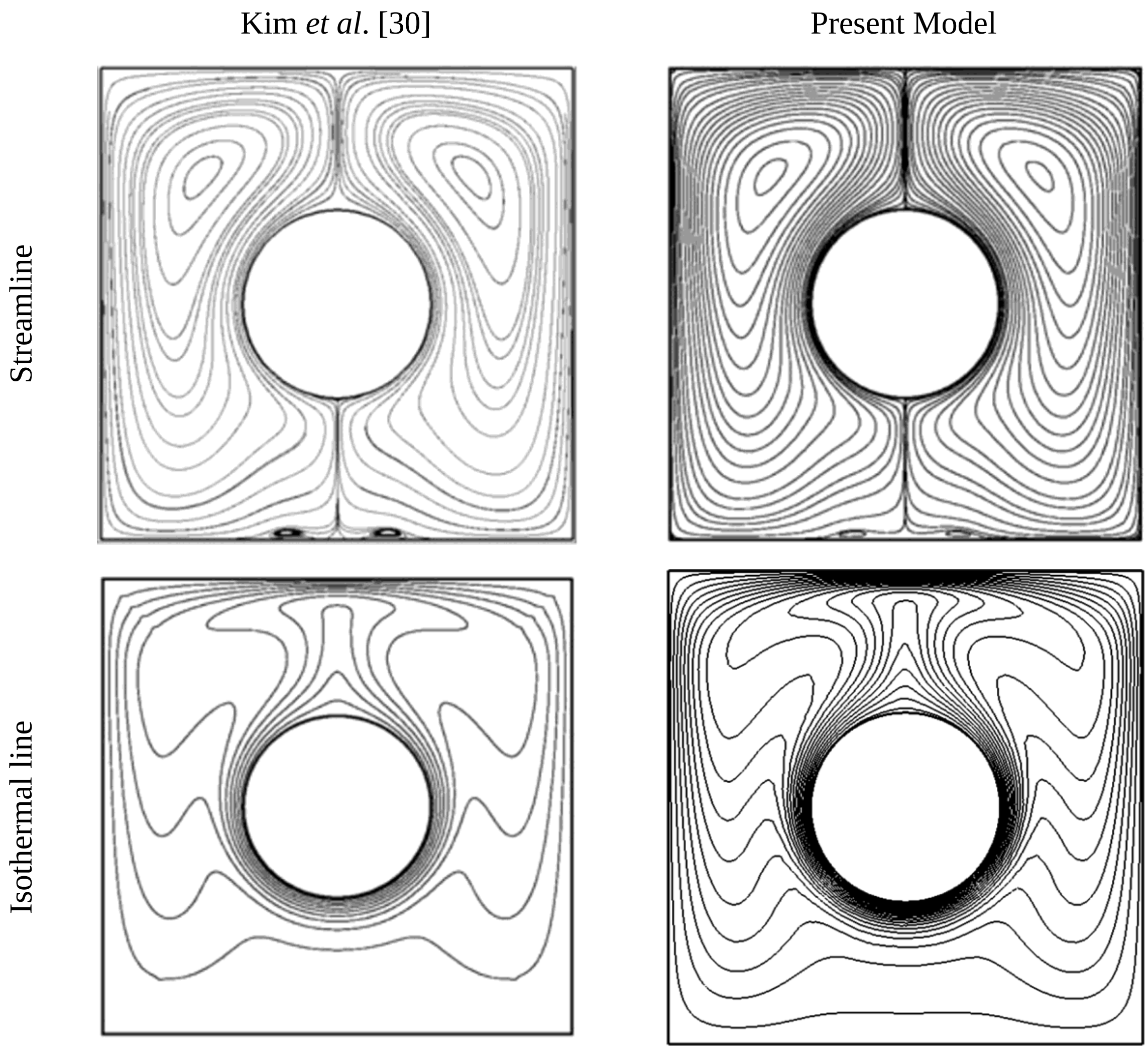

Fig 7: Comparison of streamline and isotherm line with Kim et al. [2] and present model.

| | $Nu_{avg}$ | | | |
|---|---|---|---|---|
| *Ra* | $10^3$ | $10^4$ | $10^5$ | $10^6$ |
| Present model | 5.26 | 5.89 | 7.83 | 14.71 |
| Kim *et al.* [2] | 5.01 | 5.02 | 7.78 | 14.10 |

Table 1. Comparison of average Nusselt number

## 3.8 Summary

The domain is divided into unstructured triangles. Triangular factorization method is utilized in solving the transformed linear equations from the Galerkin technique. After a brief introduction to different numerical methods to solve a set of differential equations, fundamentals of Galerkin method along with discretization of governing equations are provided. Mathematical definitions of computed parameters are presented afterwards. The following section consists of grid sensitivity test of the problem. In the last part of this chapter the details of code validation tests are provided to establish validity of numerical scheme used for present analysis.

# Chapter 4

## Results and Discussions

In this investigation, mixed convection heat transfer in a rectangular channel with a centered variable speed rotating cylinder is studied. The channel is equipped with an isoflux heater on the bottom wall and a parabolic velocity profile exists at the inlet. For the purpose of this study and for separate trials, the effect of Reynolds number and Grashof number of air inside the rectangular channel is varied while the Prandtl number is kept fixed at 0.71. In effect, the Grashof number is varied from $10^4$ to $10^5$ to investigate its impact on the isotherms and streamlines while the Reynolds number is varied between 10 and 100. The effect of variation of the Grashof number, Reynolds number and speed ratio on the flow filed, temperature field and on the heat transfer characteristics is evaluated. Here, the flow field has been analyzed in terms of the stream function whereas the temperature field has been analyzed in terms of the distribution of isothermal lines. The local and the average Nusselt numbers are used in conjunction to analyze the heat transfer characteristics at the isoflux portion of the channel. At first, the impact of Reynolds and Grashof numbers on the isotherms and flow pattern is analyzed by setting the Reynolds number to 10 and Grashof number to $10^4$. Next, the same effects are analyzed by increasing the Grashof number to $10^5$. Afterwards, these steps are repeated for Re = 50 and Re = 100. The speed ratio, however is changed from 0.5, 1.0 and 1.5 and its effects on the characteristics that are highlighted are presented. The Grashof number for the speed ratio analysis is kept fixed at $10^5$ while the Reynolds number is varied with each case. The conductivity ratio of solid to fluid is kept constant at K = 10.

## 4.1 Effect of Grashof Number and Reynolds Number

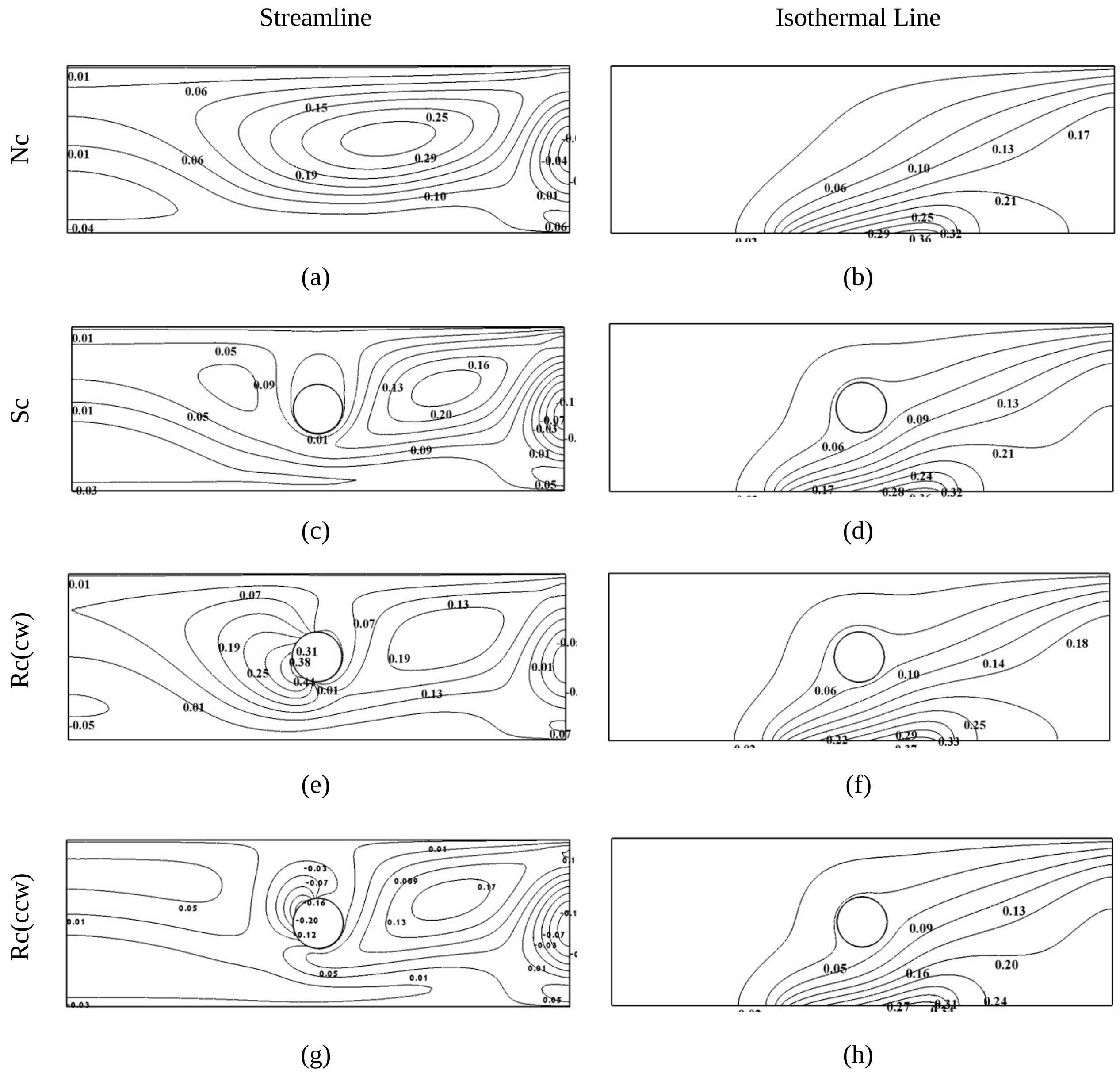

Figure 8: Streamline and Isothermal Line for *Re*=10 and $Gr=10^4$

Figure 8 illustrates the variation of Grashof number on streamline patterns and isotherms for various configurations of the cylinder namely : i) plain channel with no cylinder (nc), ii) channel with stationary cylinder(Sc), iii) channel with clockwise rotating cylinder (cw) and iv) channel with counter clockwise rotating cylinder (ccw). Here the conductivity ratio K = 10 and radius of the cylinder is 0.15H. Here the Grashof is varied from $10^4$ to $10^5$ while the Reynolds number is varied from 10 to 100.

Convection cells are generated in each case for Gr = $10^4$. For instance, in figure 8(a), a single recirculating cell develops at approximately on the middle portion of the channel. This is indicative of the dominance of the convective mode of heat transfer over other modes. It is seen that these streamlines in this case are concentrated over the lower wall, outlet and upper wall. When the Grashof number is increased to $10^5$ in figure 9, these cells increase in strength. In particular, the convection cell in Fig. 9(a) is more concentrated and stretched out in the direction of flow than in Fig. 8(a). This stretching squeezes the flow path and the convective heat transfer associated with the cell brings about a higher rate of heat transfer. It is noteworthy to observe that the fluid passing over the cylinder surface is motionless and thus a no-slip boundary condition exists in this regard. In Fig. 8(c), two vortices of varying strength are observed to form on the left and right part of the cylinder. The left vortex is weaker than the right one. However, when the Grashof number is increased to $10^5$ in fig. 9(c), both vortices increase in strength which is observed by the increase in the density of streamlines. In Fig. 8(g), a vortex is generated on the left side of the cylinder but prior to its completion, it is seen to intersect with the cylinder. In addition a secondary vortex develops on the right portion of the cylinder but it is of lower strength. As the Grashof number is increased in Fig. 9(g), this incomplete vortex becomes more symmetric about the horizontal center line of the cavity, leading to an increase in its intensity and concentration.

Streamline | Isothermal line

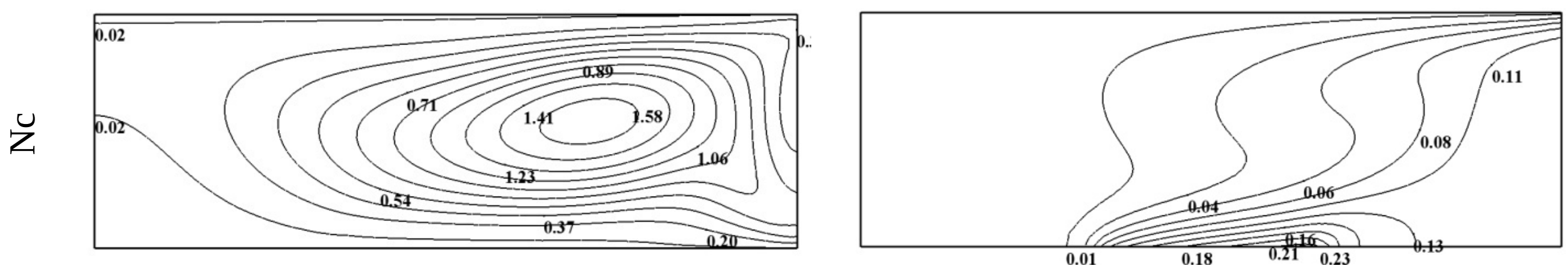

(a) (b)

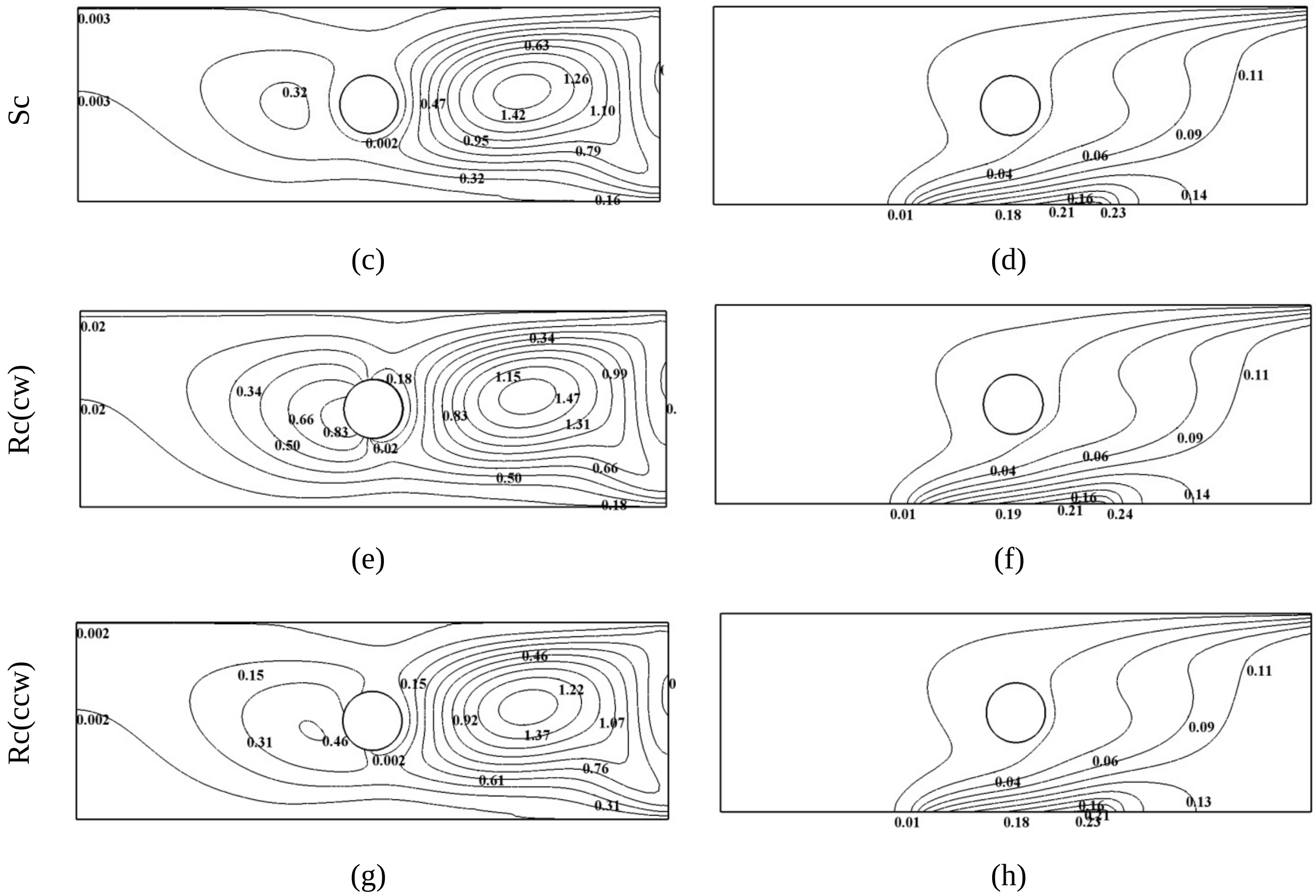

Figure 9: Streamline and isothermal line at *Re* = 10 and *Gr* = $10^5$

From Fig. 8(e) we see that, the recirculating cell to the right of the cylinder is more concentrated than in Fig. 9(e). This reduces the boundary layer thickness and effectively increases the heat transfer rate to the right of the cylinder. Comparing fig. 8(a) to 8(c), 8(e) and 8(g), it is evident that the flow field is dominated by the presence of the cylinder and its distortion is dependent of the direction of rotation of the cylinder. Figures 8(b), 8(d), 8(f), 8(h) illustrate the effect of the Grashof number on the temperature contour for $Re$ = 10 and $Pr$ = 0.71. The dense isothermal lines packing over the heater surface in all diagrams indicate a higher rate of heat transfer around this region. As the Grashof number is increased to $10^5$ in figure 9(b), 9(d) 9(f), and 9(h) the boundary layer thickness falls and these thermal cells experience a rise in their concentration of isotherm lines which leads to a consequent rise in the natural convection effect. The presence of a stationary cylinder distorts the isothermal lines distribution in Fig. 8(d) and a rotating cylinder shifts these lines in the direction of rotation of the obstacle itself. In the region to the right of the isoflux heater and at the exit, the isothermal lines intersect the cavity perpendicularly. Here heat transfer mainly occurs by conduction. For Fig. 8(d), (f) and (h), the increase in Grashof number, cause the isothermal lines to the left and right of the cylinder to widen and become more spaced out, which results in a lower rate of heat transfer from these areas. From the isotherm plots, it is seen that the main heat transfer takes place from the lower portion of the isoflux heater and the upper right side of the channel which is bounded by the upper wall and the outlet. The cold air transfers heat from the left portion of the heater and becomes hot. In doing so, the density decreases and it becomes lighter than the heavier cold air above the cylinder. This lighter air travels up until its motion is impeded by the cylinder. This causes the air to travel around the outer periphery of the cylinder to reach the cold upper wall. As it reaches the top, it cools down and becomes heavier which causes it to travel down to the heating element. This generates the convective cell in figures 8(a) and 9(a).

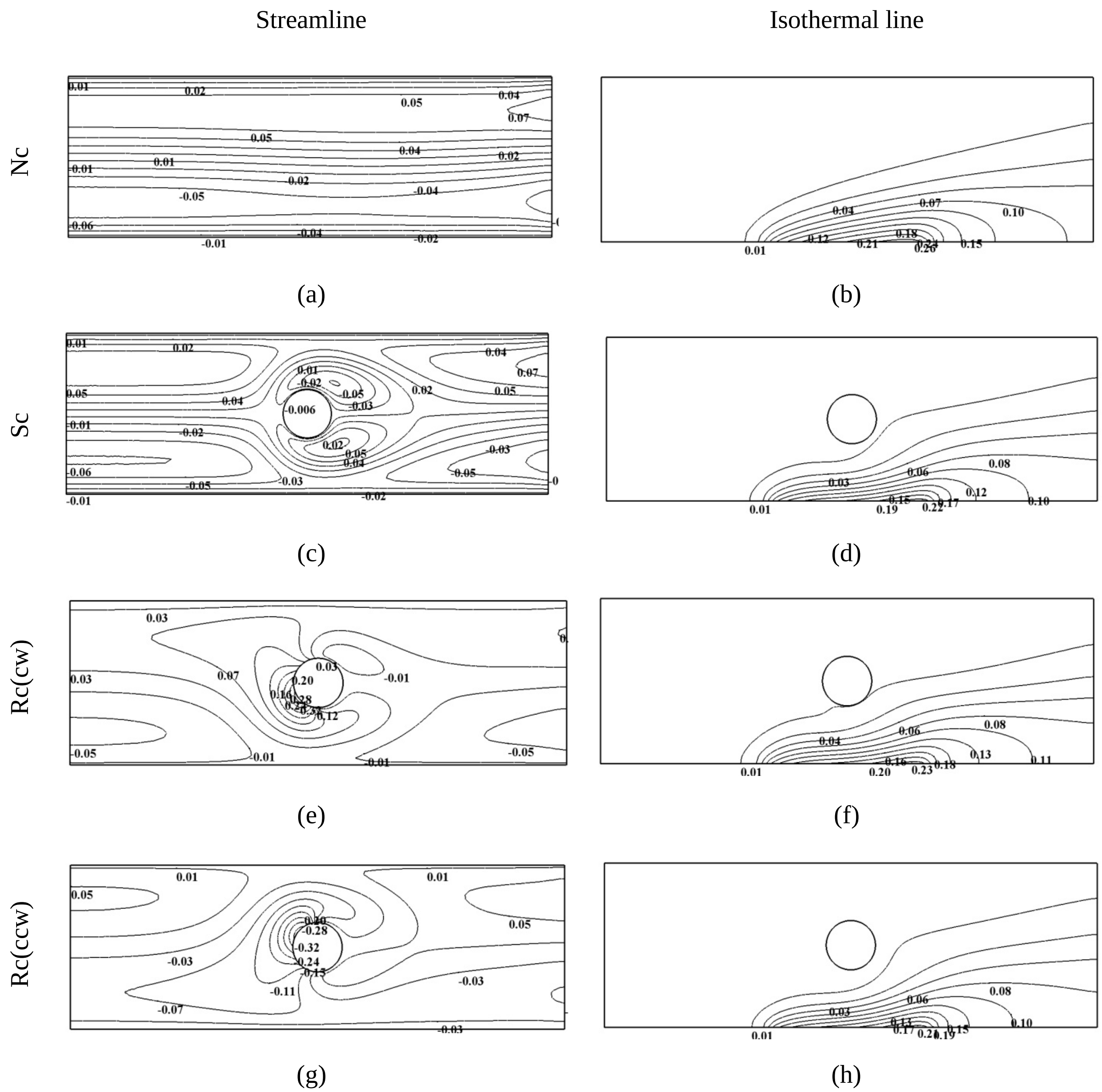

Figure 10: Stream and thermal line at $Re$=50 and $Gr$=$10^4$

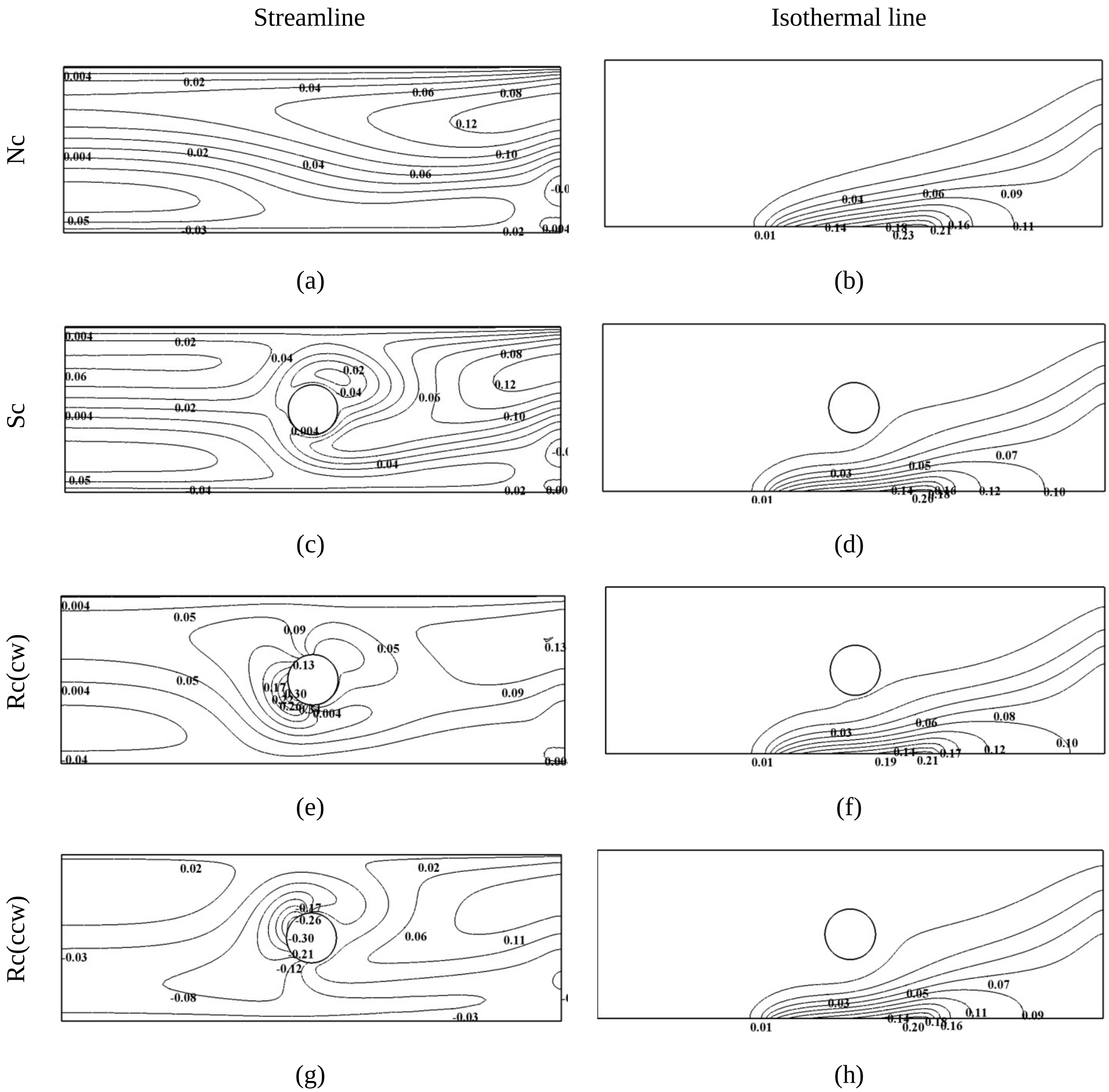

Figure 11: Stream and thermal line at *Re* = 50 and *Gr* = $10^5$

The impingement on the streamline pattern due to the Grashof number for *Re* = 50 is illustrated in Fig. 10. In the absence of a cylinder, the streamlines are seen to be almost parallel to the direction of flow and the increase in Grashof number to $10^5$in figure 11(a) has a small effect on the streamline pattern although a slight increase in the heat transfer exists due to the improvement in the convective mode of heat transfer. In Fig. 10(c), two convection cells of fairly equal strength are observed around the cylinder and with the rise of Grashof number in Fig. 11(c), the lower cell vanishes which leads to a lower rate of convective heat transfer on the lower side of the cylinder. From Fig. 10(g) it is observed that there exists a congestion of streamlines to the left of the cylinder to what appears to be an incomplete convection cell. Here, the streamlines are intersected by the cylinder before the vortex is completed. With an increase of Grashof number in this case in Fig. 11(g), the effect on the streamline pattern seems to be negligible. A similar observation can be made when comparing Fig. 10(e) and 11(e), the partial vortex in this case is generated on the lower left portion of the cylinder and just like the counter-clockwise case, the increase in Grashof number seems to have little effect. The effect of increased Reynolds number on the distribution of isothermal lines is demonstrated in figures 10 and 11. From these plots it is evident that when the Reynolds number is increased to 50, the increase in fluid inertia dominates the isothermal line distribution and thus the heat transfer characteristics. Isothermal lines emanating from the heater surface are stretched towards the direction of flow and intersect the outlet in a perpendicular manner. Thus in this region of the outlet the heat transfer becomes conduction dominated. These thermal plumes become increasingly long and stretched further in the flow direction with an increase of Grashof number in figure 11. The presence of the cylinder in this case minutely distorts the isothermal line distribution when compared to previous cases.

As the Grashof number is increased for various cases in figure 11, the thermal boundary layer over the heater surface shrinks and thus this leads to higher rates of heat transfer. Moreover the isothermal lines are seen to be concentrated in the region between the lower part of the cylinder, the bottom wall and the exit. In figures 10 and 11 the isotherms are more spaced out at the rear end of the channel, which denotes lower rates of heat transfer. By increasing the Grashof number to $10^5$, small changes are observed in the temperature field since the flow is now dominated by larger fluid inertia.

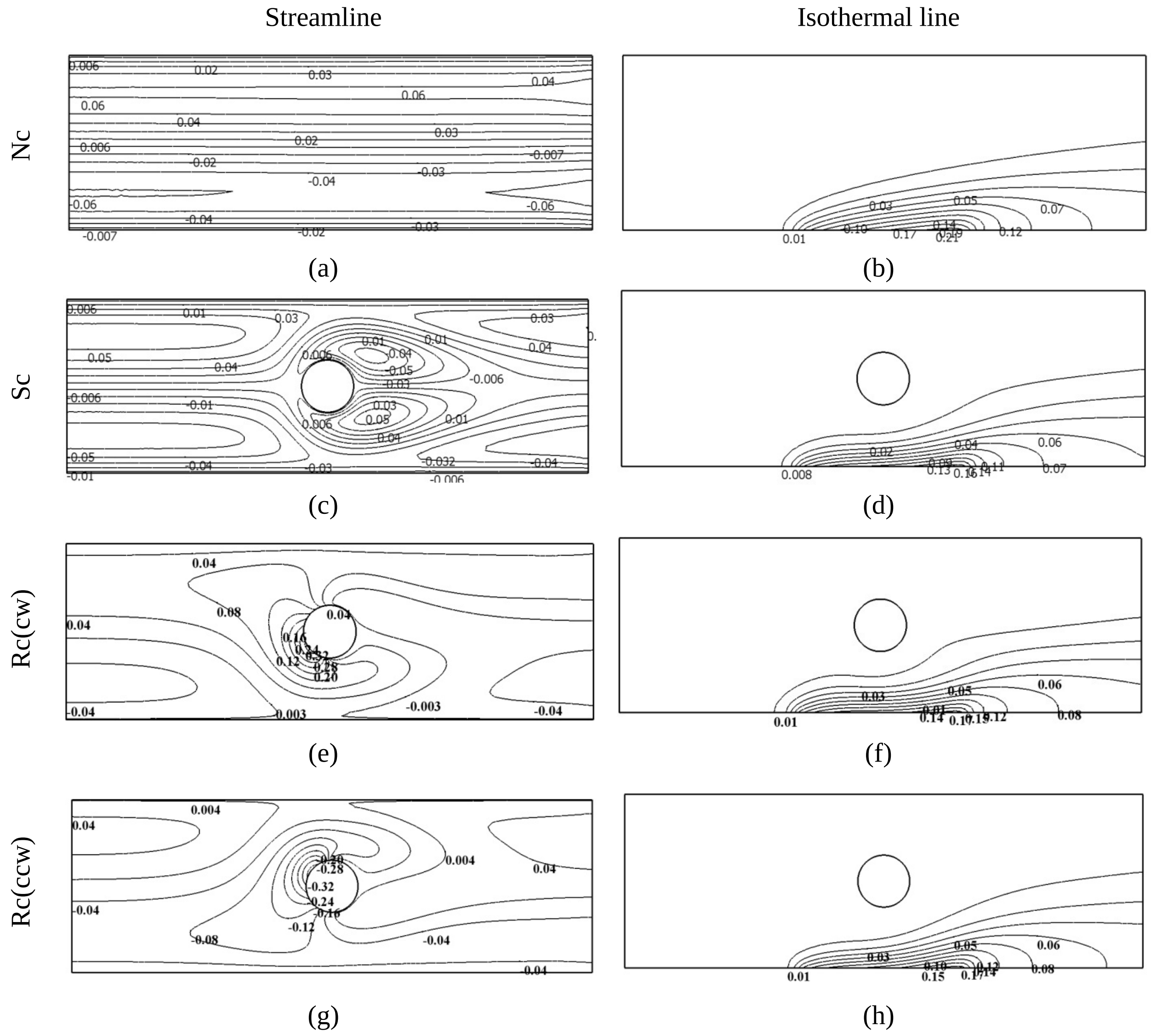

Figure 12: Stream and isothermal line at $Re$ = 100 and $Gr$ = $10^4$

Figure 12 demonstrates the effect of *Re* = 100 and Gr = $10^4$ on the isothermal lines and streamlines. This further rise in fluid inertia causes the dual vortices in figure 12(c) to be even more symmetric and elongated. This results in a slight increase in the heat transfer around the cylinder. Also these vortices are less intercepted by the cylinder's outer periphery and are thus more complete in shape and form. From the temperature filed plots in figure 12, it is observed that the isothermal lines are even more compressed against the heater and the bottom wall and are thus more concentrated. This leads to a slight decrease in the boundary layer thickness in comparison to temperature fields with a lower Reynolds number. And as a result, this leads to small increase in the rate of heat transfer.

## 4.2 Effect of Cylinder Speed Ratio

To illustrate the effect of speed ratio on the streamlines and isothermal lines, K is set to 10 and the Grashof number is fixed at $10^5$. The Prandtl number of air inside the cavity is set to 0.71. The speed ratio is varied as 0.5, 1.0 and 1.5. In figure 13, the Reynolds number is kept low at 10 and the impact of the speed ratio on two configurations (clockwise and counter-clockwise) are demonstrated.

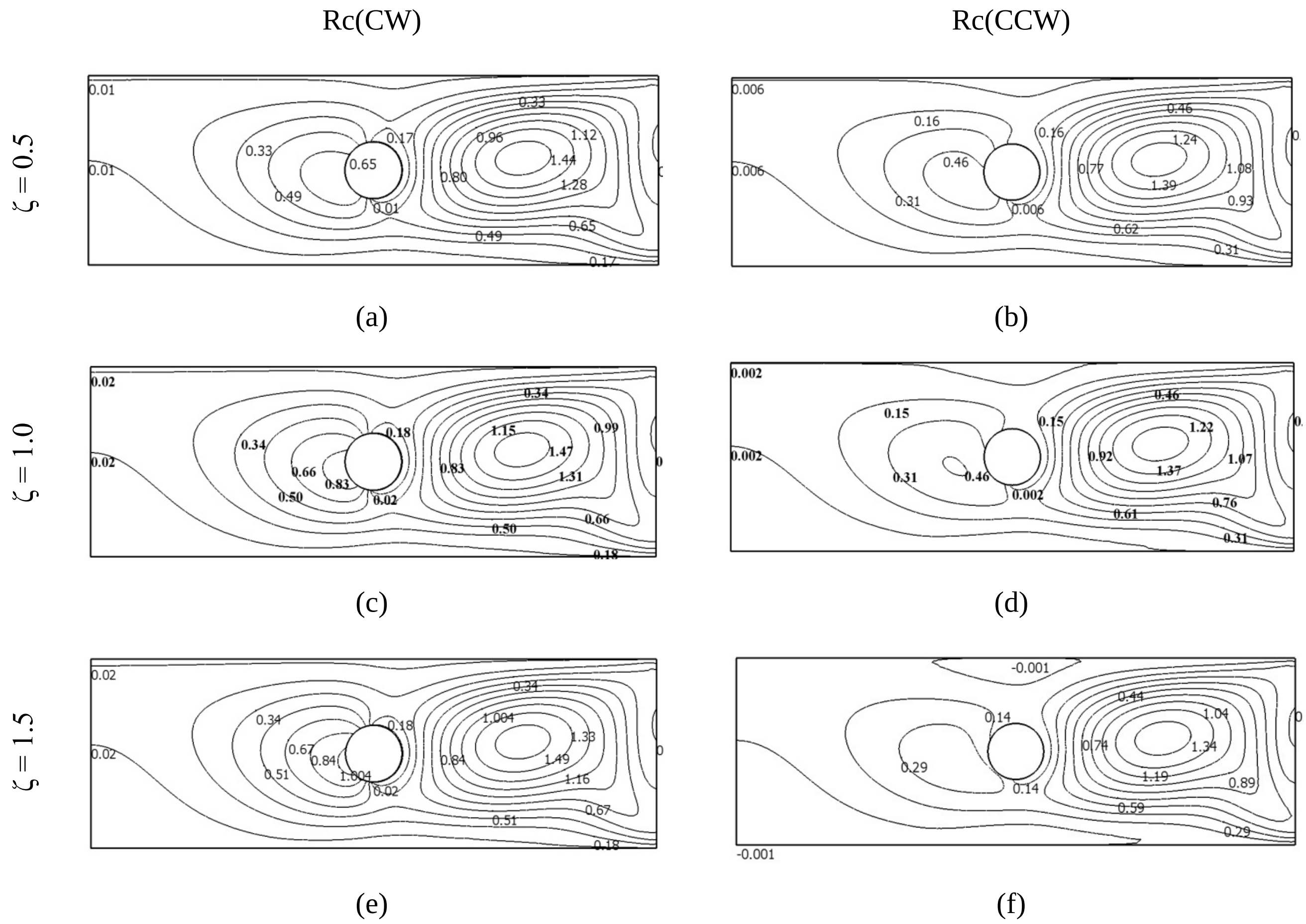

Fig 13: Effect of ζ on streamline at Re = 10 and Gr = $10^5$

In the first case in fig 13(a), the 2 vortices are seen be part of a larger recirculating cell with the cylinder rotating clockwise. The left vortex is of lower strength and thus weak compared to the vortex on the right which is stronger and hence more concentrated. This high degree of concentration reflects higher rates of heat transfer from the rear of the cylinder. The left vortex in figure 13(a) is part of an incomplete recirculating cell and the streamlines are intercepted by the cylinder surface whereas the right vortex is complete. The streamlines on the right are compressed on the upper and bottom walls which reflect higher rates of heat transfer from these areas. A small streamline is also seen to originate at the bottom of the cylinder which surrounds the rear portion. Comparing this fig 13(b), where the cylinder is rotating in the counter-clockwise direction, the vortices or eddies are narrower and division of the vortices takes place. Since the eddy on the right hand side of cylinder becomes more squeezed, the streamlines become more congested and thus a higher rate of convective heat transfer results. It is also interesting to note that the vortex on the left side of the cylinder becomes more spaced out and thus the distance between adjacent streamlines fall. Thus it becomes evident that the direction of rotation of the cylinder plays an important role in the flow field. When the speed ratio is increased to 1.0, the rotational speed of the cylinder gradually takes priority over the fluid inertia. Both vortices on the left side of the cylinder becomes thinner and narrower in figure 13(c), while it seen that only the vortex on the right portion becomes stronger in figure (d). This implies a shift in the degree of heat transfer from the left portion to the right. Further increase in the speed ratio to 1.5 seems to have little effect on the streamline patterns for both cases.

Figure 14 demonstrates the effect on the streamline distribution when the Reynolds number is increased to 50. As the Reynolds number is increased, it becomes quite apparent that large fluid inertia controls the streamline distribution.It can be observed that partial convection cells are generated in each case over the top and bottom half of the cylinder. In all cases where incomplete vortices are formed, the streamlines are intersected at the cylinder surface before the re-circulating cells could be completed. Here, it is evident that the direction of rotation of the cylinder plays an important role in controlling the streamline distribution around it.

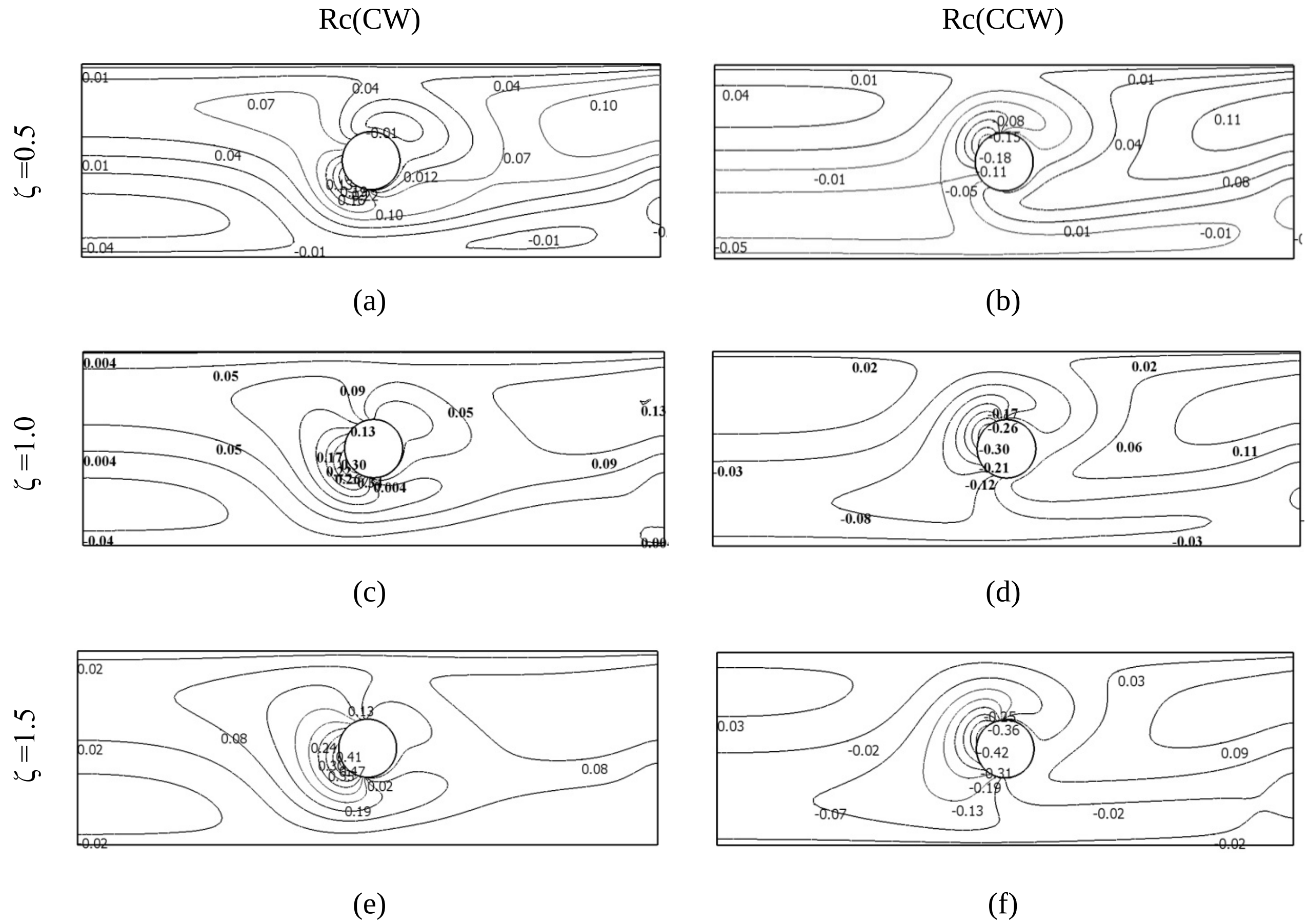

Figure 14: Effect of ζ on streamline at Re=50 and *Gr* = $10^5$

For instance in 14(a), two vortices are formed over the top and bottom half of the cylinder. The lower vortex is stronger and positioned more on the front half while the upper vortex is weaker and is placed a bit downstream over the top half.A small recirculating zone appears on the lower right of the channel. This suggests higher rates of heat transfer from the cylinder surface and from the bottom from the bottom of the channel, over the heater surface. The streamlines appear to be more elongated and aligned with the flow direction. For instance the flow path to the top right of 14(b) is more compressed than it was in 14(a). However the counter clockwise rotation (ccw) of the cylinder in 14(b) places the stronger vortex in the upper front part of the cylinder towards the direction of flow and that the distance between adjacent streamlines in this region is lower compared to that of the corresponding cell generated in figure 14(a). As a result the streamlines in this section are more densely packed which translates to a higher rate of convective heat transfer. As the speed ratio is increased to 1.5, the rotational effect of the cylinder takes dominance and the dual cell in 14(a) changes to a stronger lower cell with a weaker upper cell over the cylinder surface. Thus the streamlines around the lower cell becomes more concentrated. The lower right vortex is also seen to disappear as the speed ratio is increased, indicating lower rates of heat transfer in the region of the lower outlet and bottom wall. In comparison to figure 14(d), the single partial cell on the front seems to grow in strength. As the speed ratio is further increased to 1.5, all consecutive vortices increase in strength and thus it leads to slightly higher rates of heat transfer.

As the Reynolds number is further increased to 100 in fig. 15, forced convection effects dominate over natural convection effects. The streamlines in all figures are more straightened out and aligned with the flow direction. In fig 15(a), the top vortex becomes very stretched and elongated, which cause the flow around the top surface to be more squeezed. Similarly the partial vortex on the bottom surface of the cylinder decreases in size at the core and becomes increasingly concentrated. Comparing this to fig 15(b), the secondary lower vortex is shifted further downstream, indicating higher rates of heat transfer from this region while the top partial vortex only seems to increase in strength. For the other configuration, streamline density in regions away from the cylinder increase and the streamlines become more flat as a result of high fluid inertia. Consecutively, the streamline density around the cylinder in cases for ζ = 1.0 and ζ = 1.5 slightly increase and result in enhanced rates of heat transfer with clearly visible partial recirculating cells.

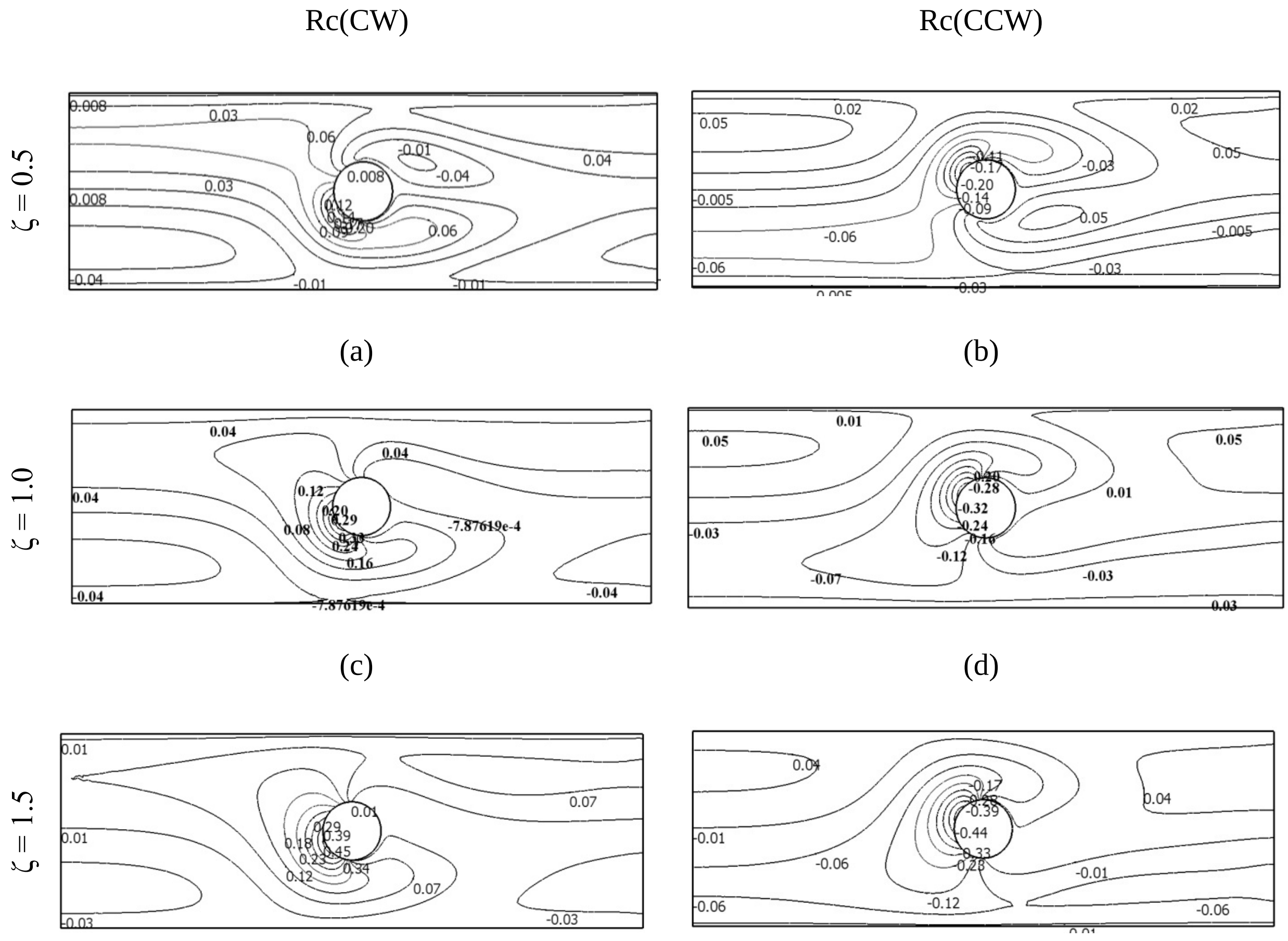

(e) (f)

Figure 15: Effect of ζ on streamline at *Re* = 100 and *Gr* = $10^5$

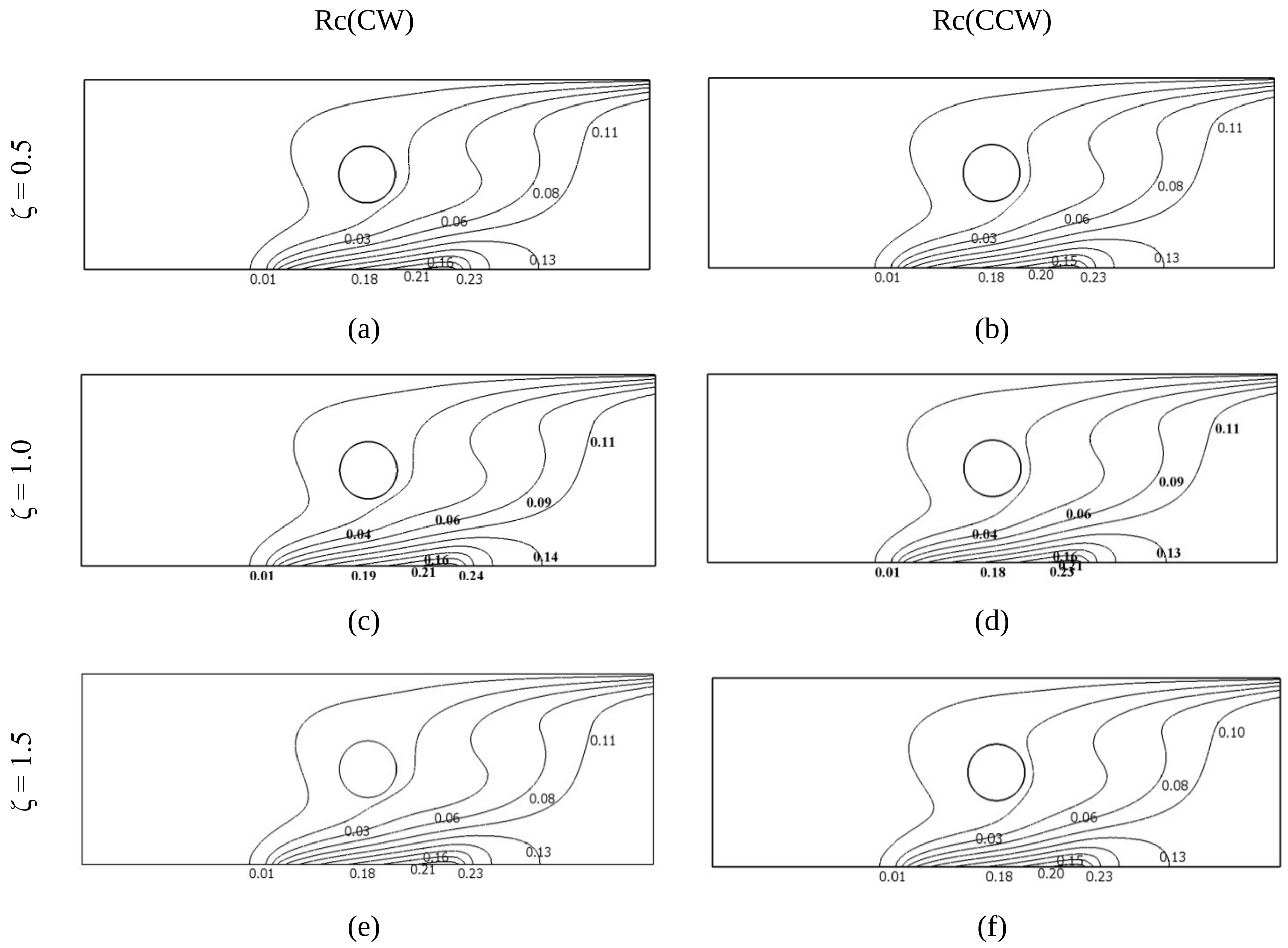

Figure 16: Effect of ζ on streamline at *Re* = 10 and *Gr* = $10^5$

Figure 16 demonstrates the effect of effect of speed ratio on the isothermal line distribution for Reynolds number set to 10 and Grashof number set to $10^5$ . Here the Prandtl number of air inside the channel is 0.71 and the speed ratio is varied as 0.5, 1.0 and 1.5. The conductivity ratio of solid to fluid is unchanged at 10 while the radius is kept fixed at 0.15H. The effects are demonstrated for the clockwise and counter-clockwise rotation cases. It is seen that for every configuration, the isothermal lines are flat over most of the surface of the isoflux heater. Also the isothermal lines are very congested over the heater surface and thus the isothermal line density is higher. As a result, the boundary layer thickness is small and this leads to higher rates of heat transfer. since the isothermal lines are very concentrated over the heating element, it can be inferred that heat is mainly transferred by convection in these cases. For all configurations it is observed that the isotherms demonstrate a sharp change in their gradient with some distance away form the heater and thus rise almost vertically around the cylinder. this indicates that the mode of heat transfer in these regions are largely due to natural convection. The isothermal line gradient around the cylinder is steep conpared to other places in the fluid where the slope of of the isotherms are low. This indicates continuity prevailing in the solid-fluid interface. For all cases, where the iosthermal lines are close together, the temperature gradients in these regions are consequtively large and this results in higher heat transfer rates. The isotherms are also oberved to be more spaced out in the middle regions that are downstream of the cylinder and thus the temperature graident in this region is lower and heat transfer rates are lower. Also fewer isothermal lines are observed on the left side of the cylinder compared to the right. This indicates that heat transfer is mainly concentrated in the regions behind the cylinder and above the heating element. As the spped ratio is increaded to 1.0 and 1.5 it seems to have litte effect on the isothermal line distribution with little changes in the gradient of the isotherms around the cylinder from case to case.

In figure 17, the Reynolds number is increased to 50 and consequtively, the fluid inertia rises. This plays an important role in shaping the temperature contour since the few isoterms on the left side of the cylinder have disappeared which shows that the heat transfer is now concentrated on the lower portion of the cylinder, the bootm wall and the outlet. In 17(a), the change in gradient of the isotherms becomes more gradual as the the Reynolds number is increased. Furthermore the isotherms that extend downstream of the cylinder have an almost constant gradient. The isothermal line around the cylinder is now less steep than before and the overall isothermal line gradient is lower than before

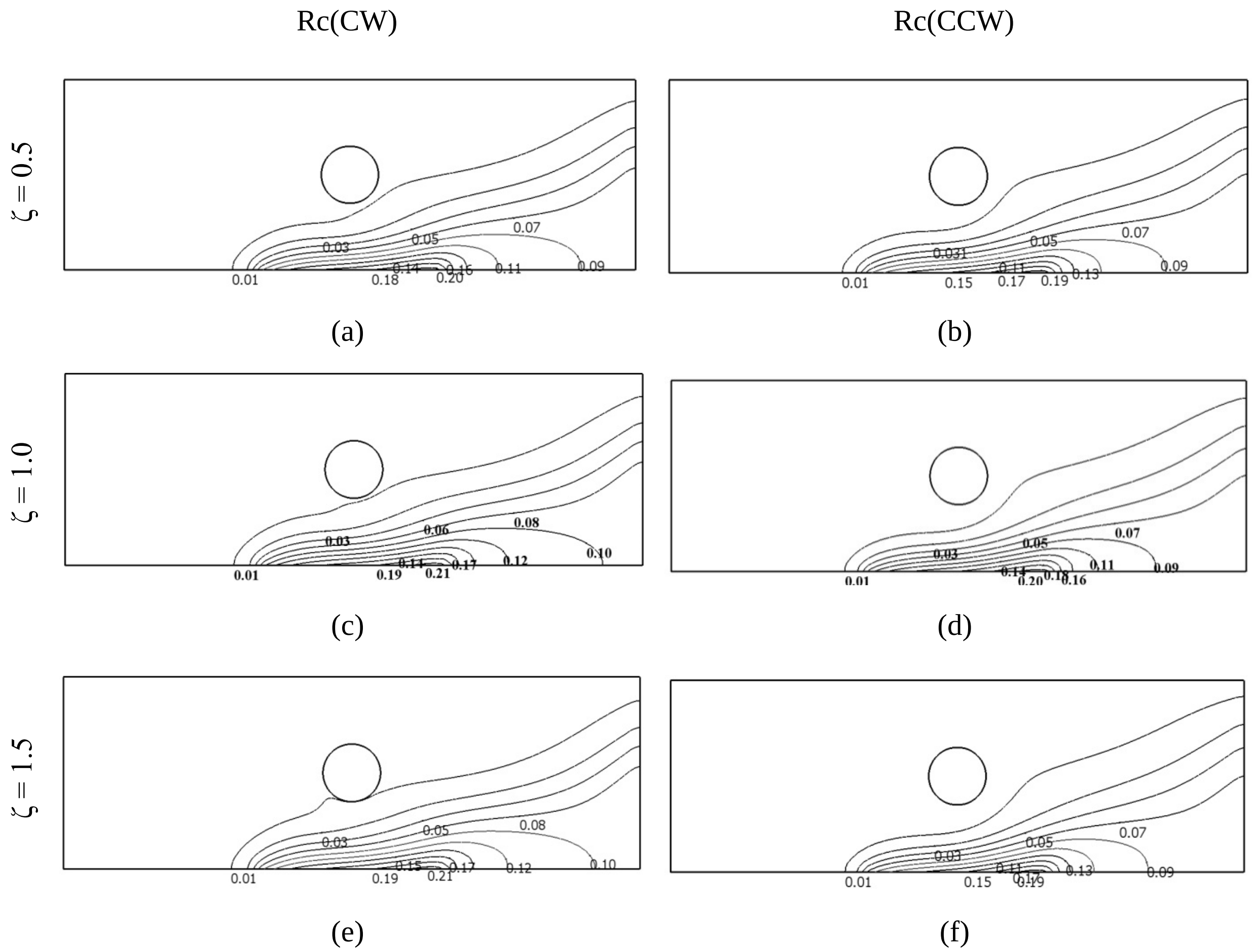

Figure 17: Effect of ζ on streamline at $Re$ = 50and $Gr = 10^5$

In fig 17(a), the isotherms are closer to the cylinder surface when compared to 17(b). As the speed ratio is increased to 1.0 in 17(c), the isothermal line approaches the cylinder while for the counter-clockwise case it recedes form the cylinder. However the slope of the isotherm in the solid region in the clockwise case is smaller than the corresponding counter-clockwise case. With a further increase in the speed ratio to 1.5, the isothermal lines in the counter-clockwise case become more condensed while those of the clockwise case become more spaced out, indicating lower rates of convective heat transfer.

In figure 18, the Reynolds number is further increased to 100 which significantly increases fluid inertia and associated effects. The fluid inertia is so high in this case that any rise in the speed ratio for different configurations has negligible impact on the temperature distribution. To understand its effect, a closer look at the local and surface averaged Nusselt number plots must be done. The isothermal lines in all cases become even more flat over the heater surface and downstream of the cylinder which causes the isotherms over the bottom surface to be even more compressed and concentrated. Subsequently, this increases the rate of convective heat transfer over the heating element and the bottom wall. Like before, an increase in the speed ratio spaces out the isotherms for the clockwise case and squeezes the isotherms for the counter-clockwise case. This leads to lower rates of heat transfer in these regions for the clockwise case and higher rates of heat transfer from these areas in the counter-clockwise case. It is interesting to note that in fig. 18(f), the isothermal line at the cylinder experiences an abrupt change in gradient from steep to almost flat inside cylinder, followed by a steep negative gradient and finally and small positive gradient at the outlet. This leads to higher rates of heat transfer in the solid region.

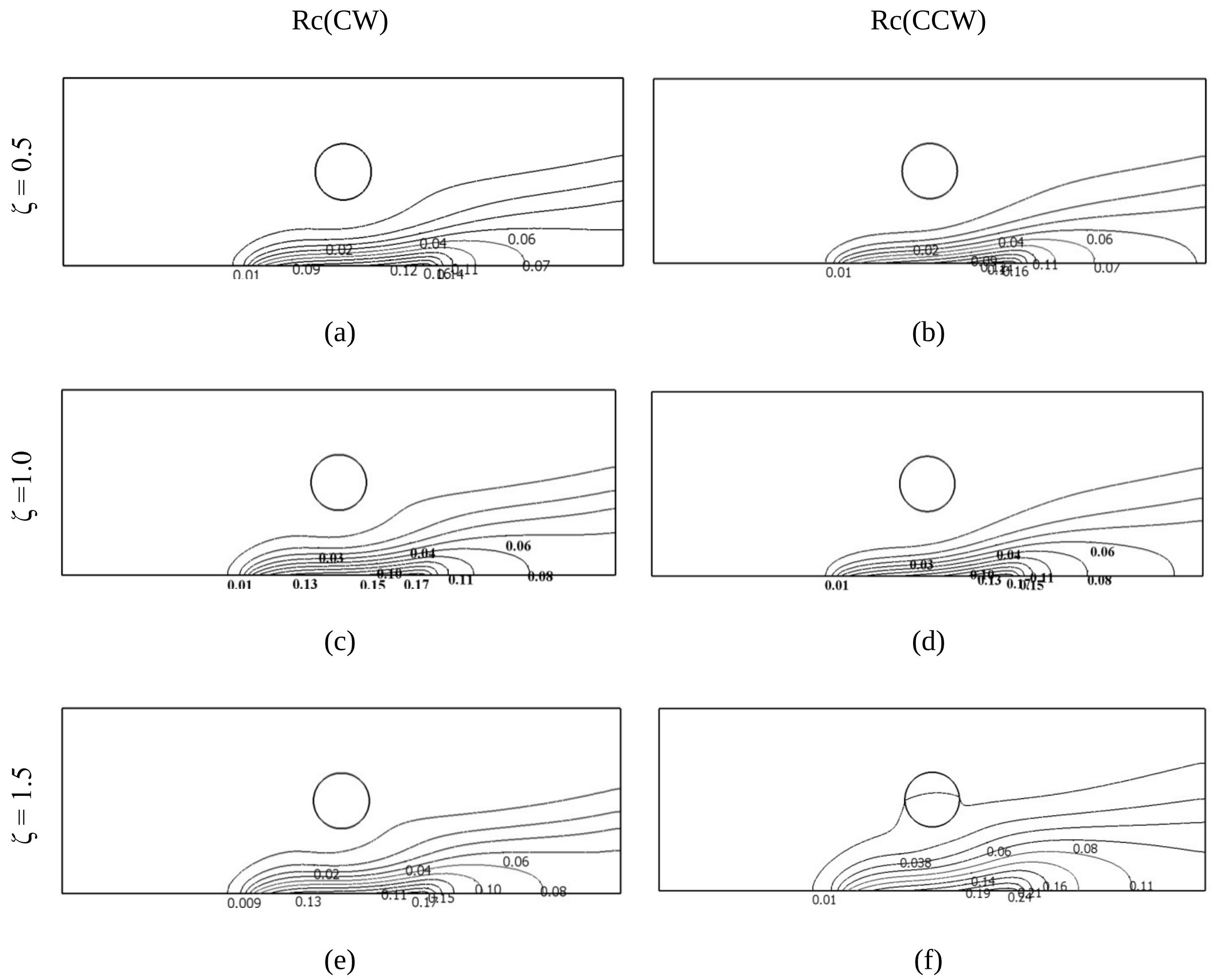

Figure 18: Effect of ζ on streamline at *Re*=100and $Gr = 10^5$

## 4.3 Effect on Heat Transfer Characteristics

In the previous sections, effects of Grashof number, Reynolds number and speed ratio on the flow field and temperature field have been analyzed. The analysis of heat transfer is based on the examining the plots of the local Nusselt number and the surface averaged Nusselt number. The variation in these number are directly correlated to the heat transfer characteristics.

### 4.3.1 Variation of $Nu_l$ and $Nu_{avg}$ with *Gr* & *Re* for various cases

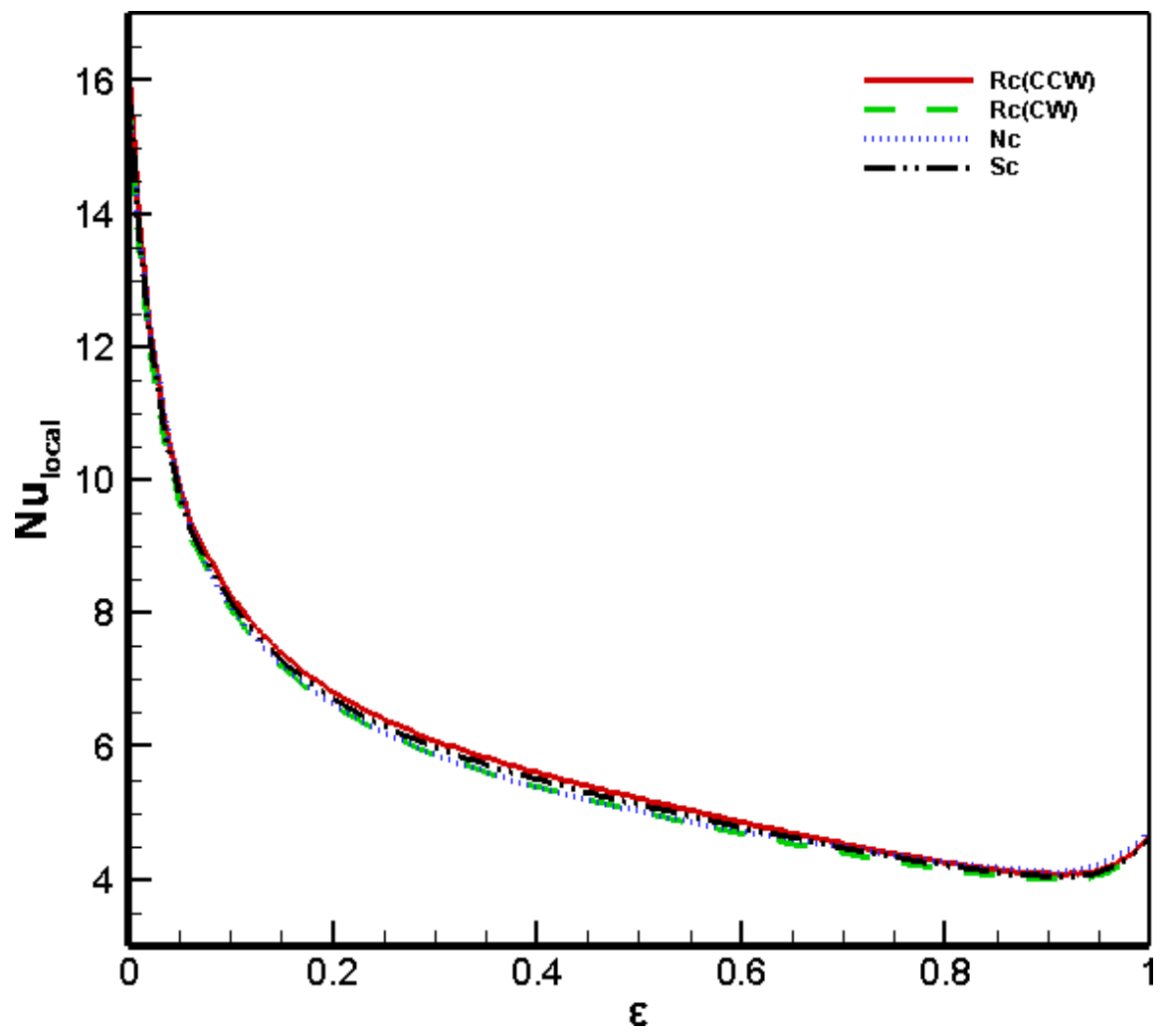

Fig. 19(a) $Nu_l$ vs heater position at $Re$ = 10

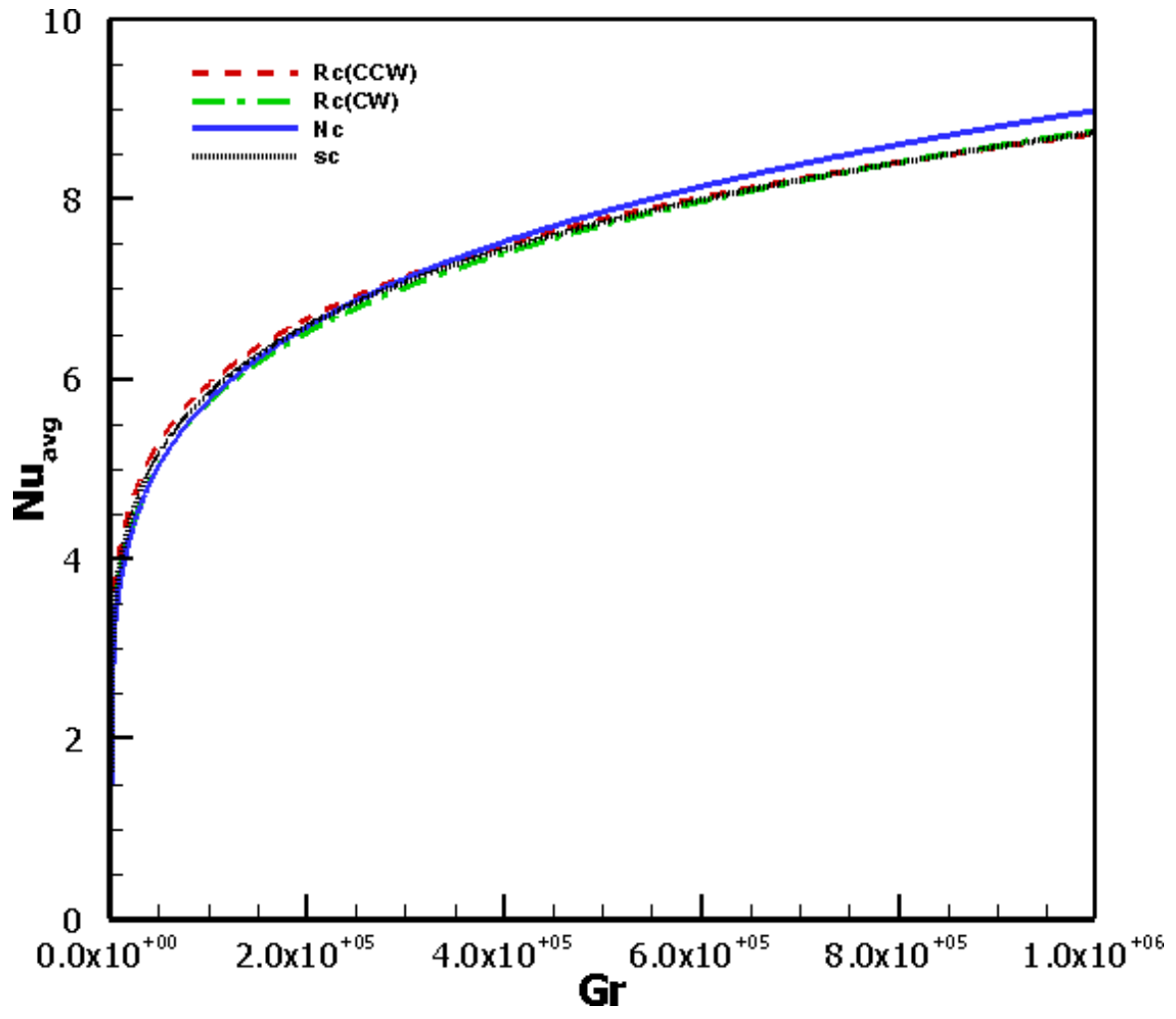

Fig. 19(b) $Nu_{avg}$ with *Gr* at $Re$ = 10

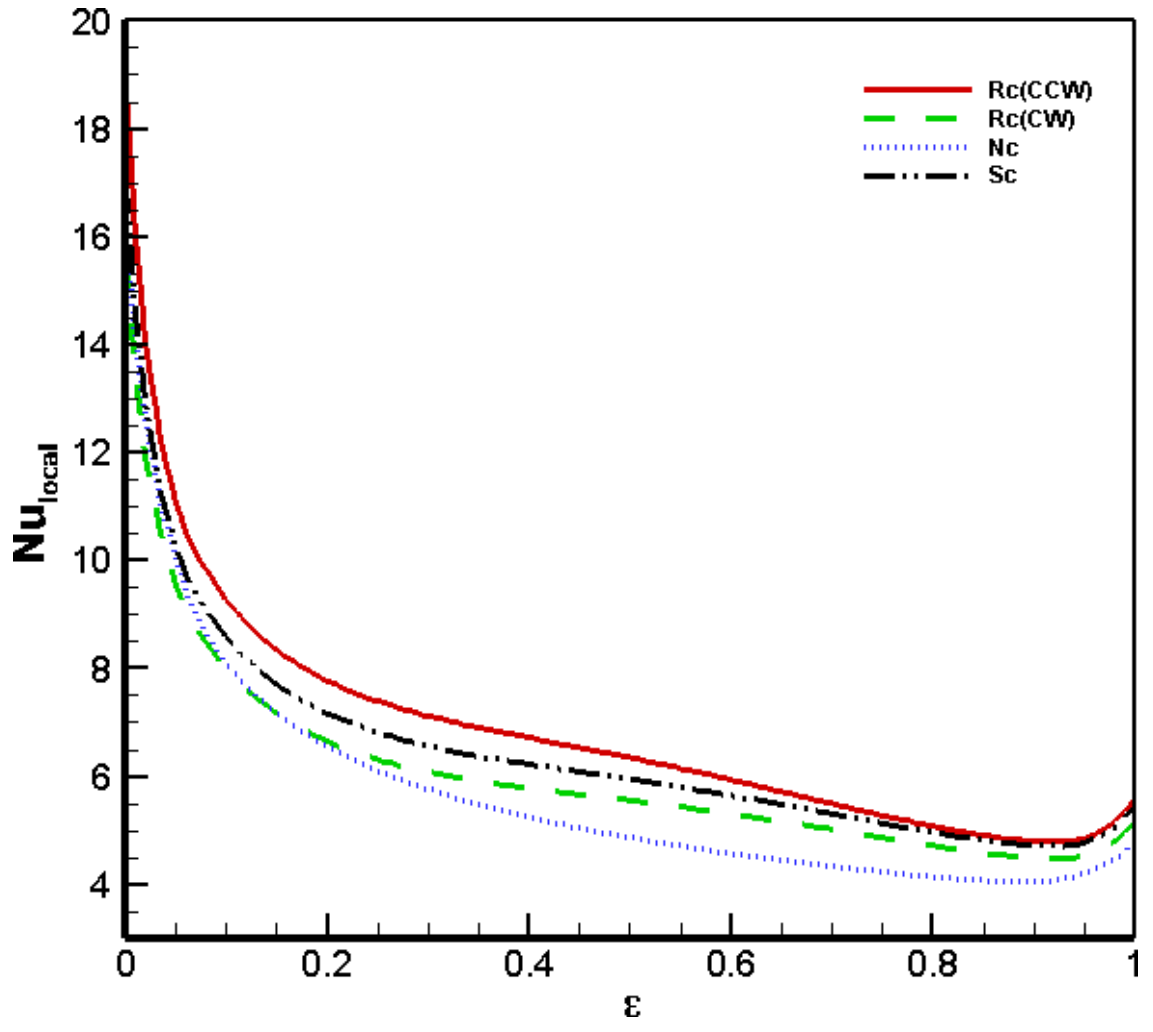

Fig. 20(a) $Nu_l$ vs heater position at $Re$ = 50

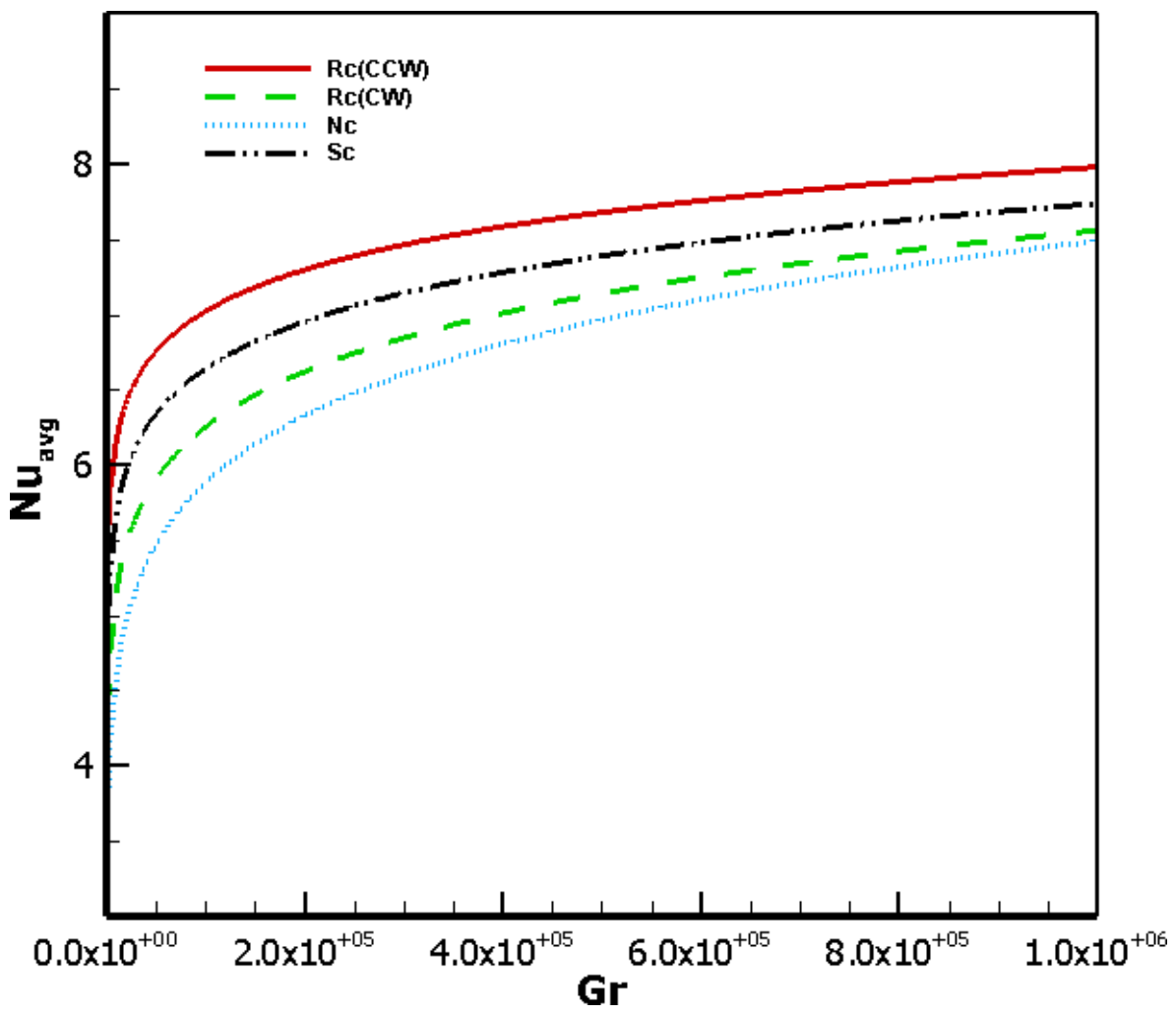

Fig. 20(b) $Nu_{avg}$ with *Gr* at $Re$ = 50

The local and average Nusselt number for the isoflux heater versus various parameters have been plotted in Fig. 19, Fig. 20 and Fig 21. Here, the change of $Nu_l$ and $Nu_{avg}$ over the arc length of the heating element and the Grashof number respectively is used to assess the heat transfer phenomena. Subsequently, the prominence of the convective mode of heat transfer over other modes is indicated by large values of the Nusselt number. Figure 20(a) shows the variation of local Nusselt number for different configurations of the rotating cylinder namely counterclockwise rotation (CCW), clockwise rotation (CW), no cylinder (NC) and stationary cylinder (SC). Here $Pr = 0.71$, $Re = 50$ and $Gr = 10^5$. The maximum Nusselt number occurs at the left side of the heating element where it peaks at $Nu_l = 19$ in Fig. 20(a). Hence this leads to an increase in heat transfer in light of the information that the boundary layer is observed to be thin in this region. Figure 19(a) demonstrates that heat transfer characteristics for all configurations are similar while in Fig. 20(a), the counterclockwise rotation of the cylinder results in the highest amount of heat transfer followed by stationary cylinder, clockwise rotation of the cylinder and no cylinder. This is inferred from the fact that the plot of $Nu_l$ versus arc length of the heater for the clockwise rotation of the cylinder is higher than those obtained for stationary cylinder, clockwise rotation of the cylinder and no cylinder. However in Fig. 20(a), the $Nu_l$ versus arc length plot for the clockwise rotation of the cylinder is seen to slightly dip under the plot for no cylinder at the left portion of the isoflux heater. For all the plots in Fig. 19(a), 20(a), the local Nusselt number peaks around the left portion of the graph where it reaches its maximum value. Afterwards, the $Nu_l$ undergoes a steep decrease followed by a shallow declining region. The $Nu_l$ plots reach a minimum value near the right hand side before rising until the end of the isoflux heater section. It is also evident that plots in Fig. 20(a) peak at higher values of the Nusselt number than its consecutive plots in Fig. 19(a) and that the Nusselt number values in Fig. 20(a) remain higher than its Fig. 19(a) counterparts for major portions of the graphs. This indicates higher rates of convective heat transfer when Reynolds number is increased from 10 to 50 between Fig. 19(a) and 20(a).

Figures 19(b) and 20(b) exhibit the variation of the surface averaged Nusselt number, $Nu_{avg}$ with Grashof number for different configurations of the cylinder which also includes its absence. From these figures, it can be inferred that $Nu_{avg}$ demonstrates a strong relationship with the Grashof number. In every instance, the $Nu_{avg}$ is observed to rise with an increase in the Grashof number. At first, $Nu_{avg}$ shows a sharp increase followed by almost linear increasing trend until $Gr = 10^6$. In all the plots of Fig. 20(b), the $Nu_{avg}$ versus $Gr$ plot for counterclockwise rotation of the cylinder is above the plot obtained from that of the stationary cylinder followed by clockwise rotation of the cylinder and finally no cylinder. For all cases, since the Reynolds number is fixed, an increase in the Grashof numbers results in an increase in the Richardson number. A Richardson number value less than 10 implies heat transfer by forced convection while a Richardson number value over 10 results in natural convection. Thus it can be inferred that when the Grashof number is increased, an increase in the Richardson number in our study makes heat transfer by natural convection prominent.

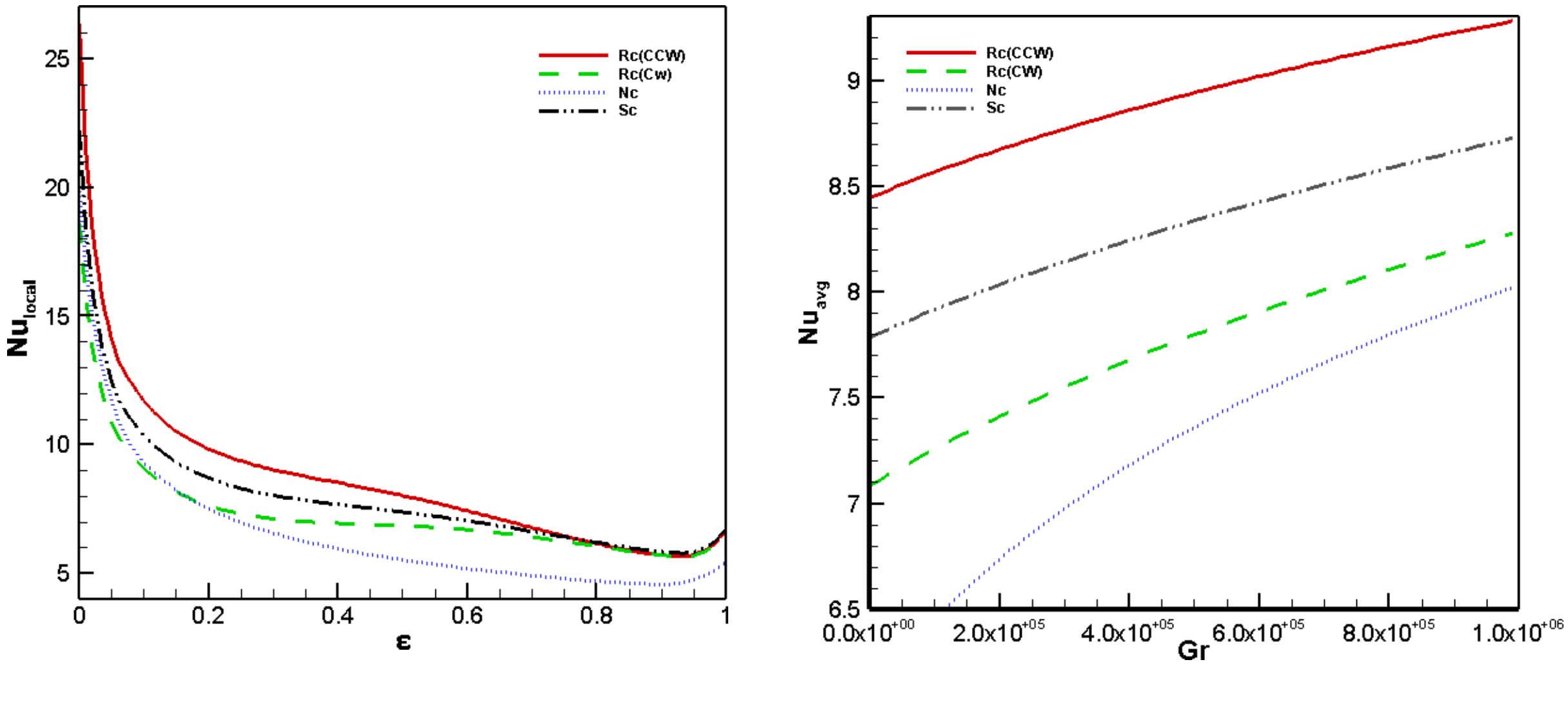

Fig. 21(a) $Nu_l$ vs heater position at $Re = 100$ Fig. 21(b) $Nu_{avg}$ with $Gr$ at $Re = 100$

In figure 21 where the Reynolds number has been increased to 100, the average Nusselt number plots become increasingly linear and similar enhancement in heat transfer rates for the local Nusselt number plots is observed. The local Nusselt number also peaks at over 25.

### 4.3.2 Variation of $Nu_l$ and $Nu_{avg}$ with Speed Ratio

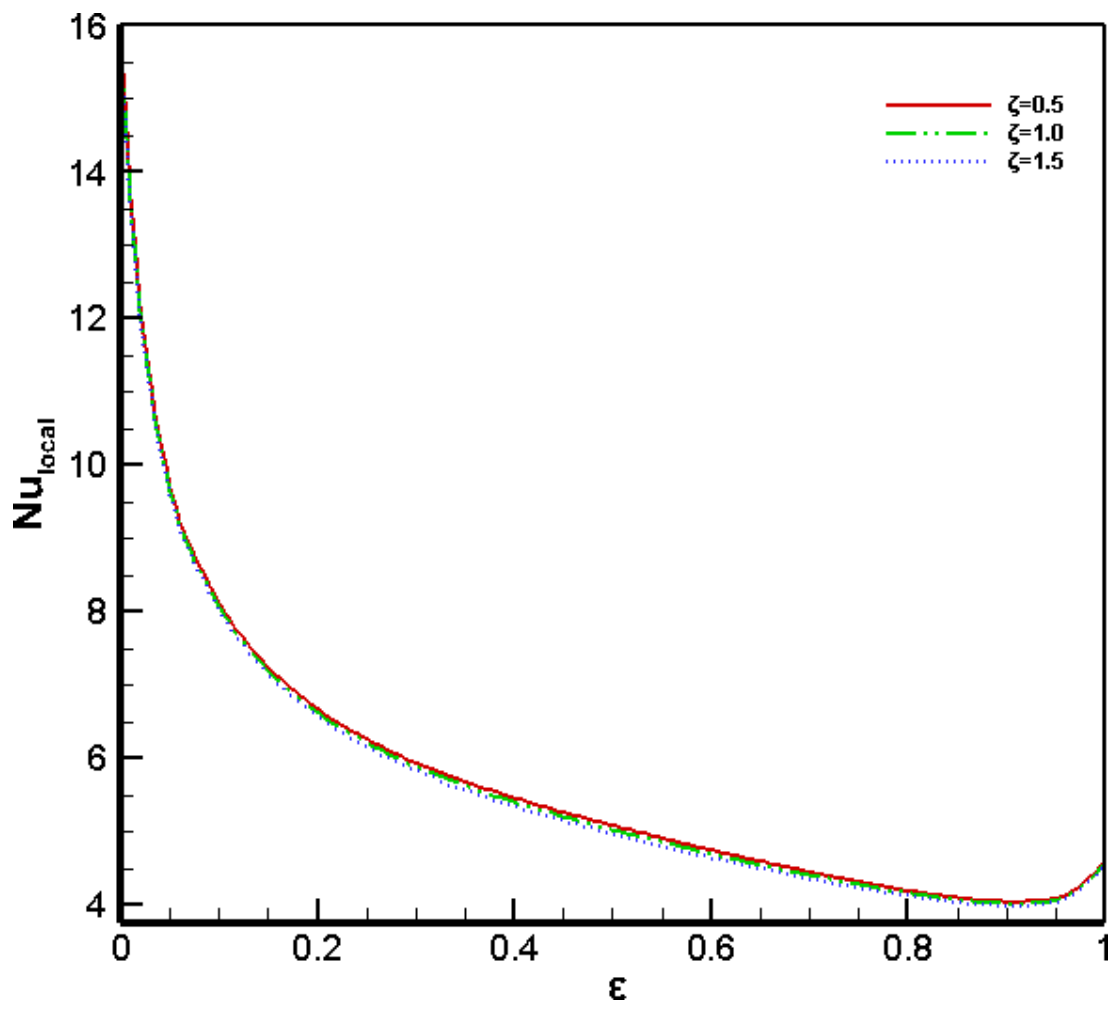

Figure 22(a). $Nu_l$ vs heater position at $Re = 10$ (cw)

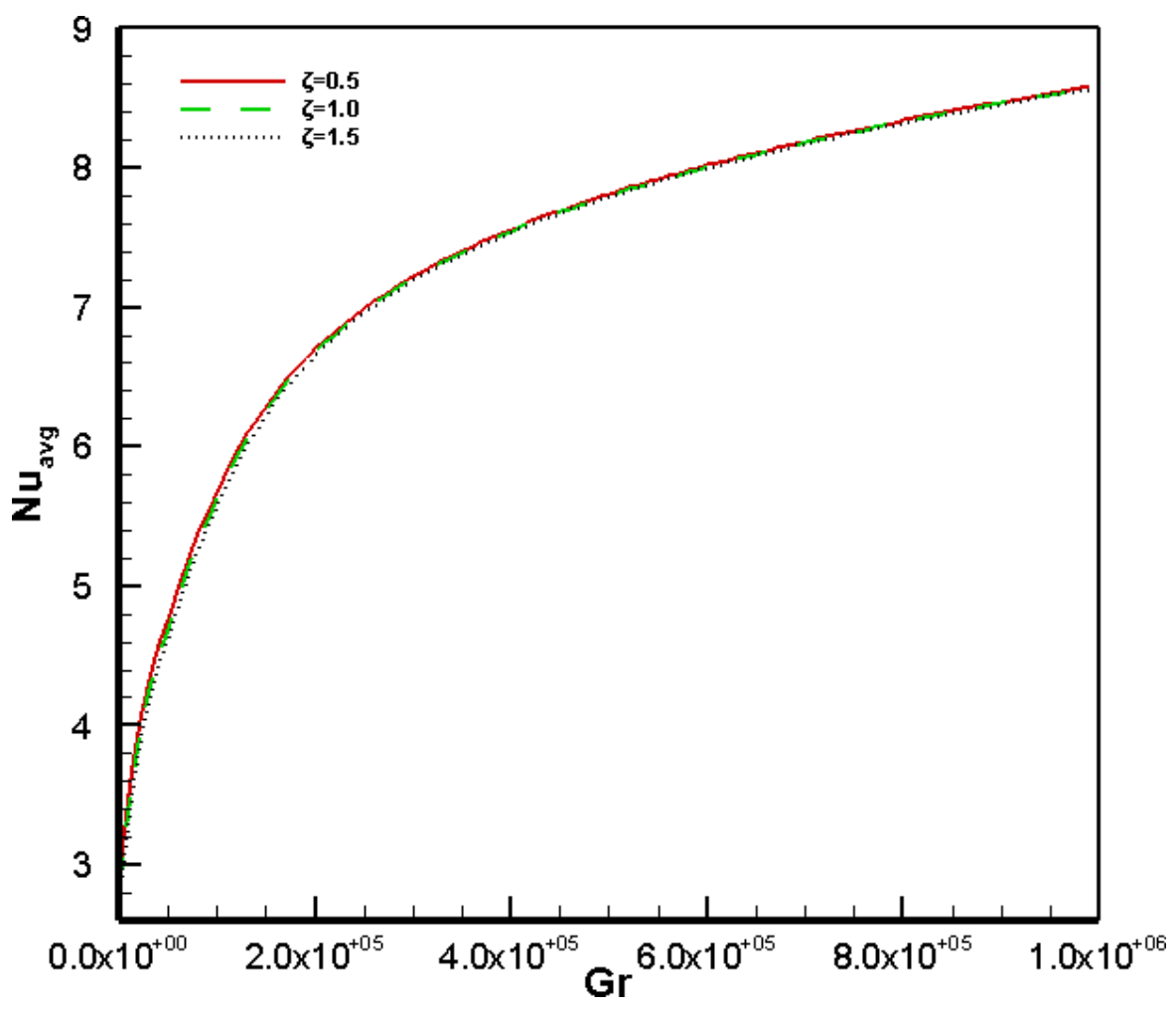

Figure 22(b). $Nu_{avg}$ vs $Gr$ at $Re = 10$ (cw)

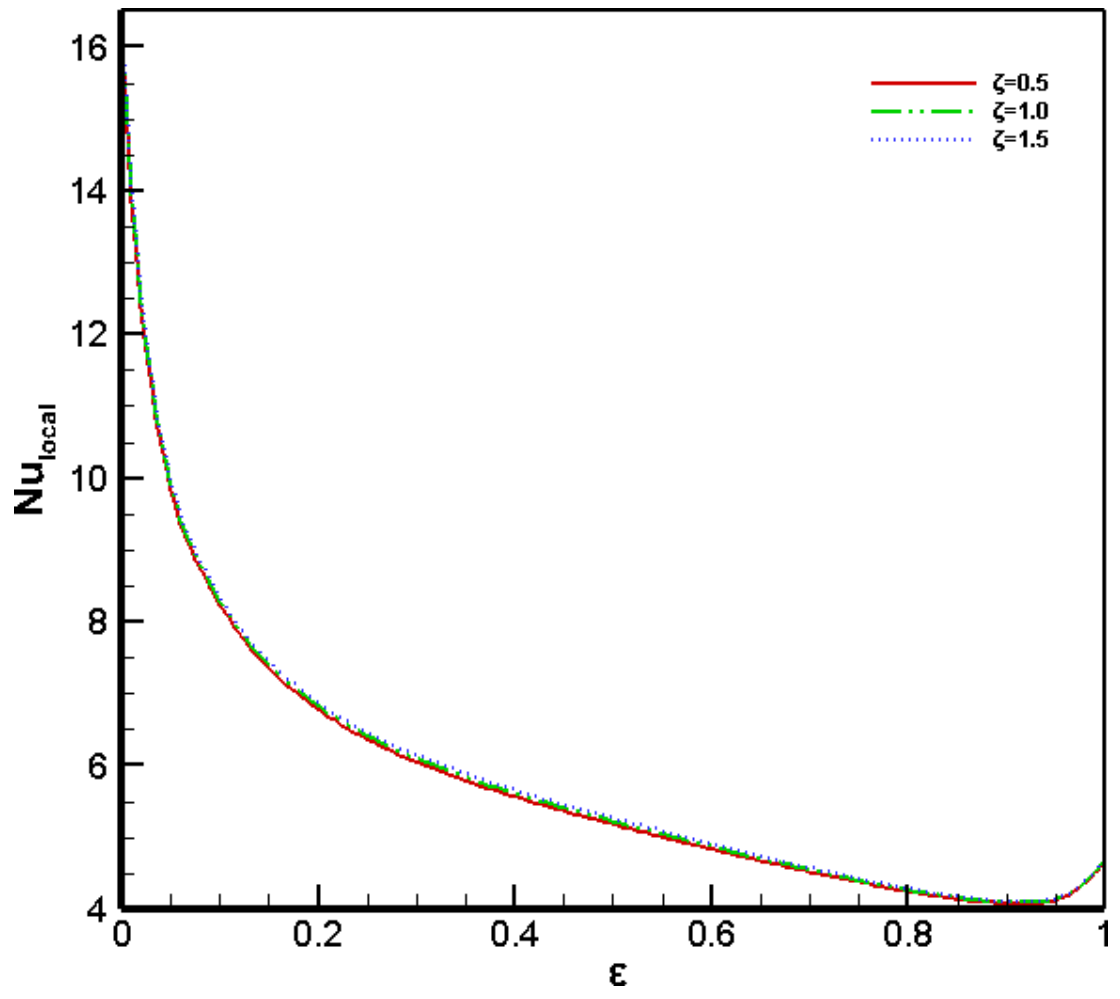

Figure 23(a). $Nu_l$ vs heater position at $Re = 10$ (ccw)

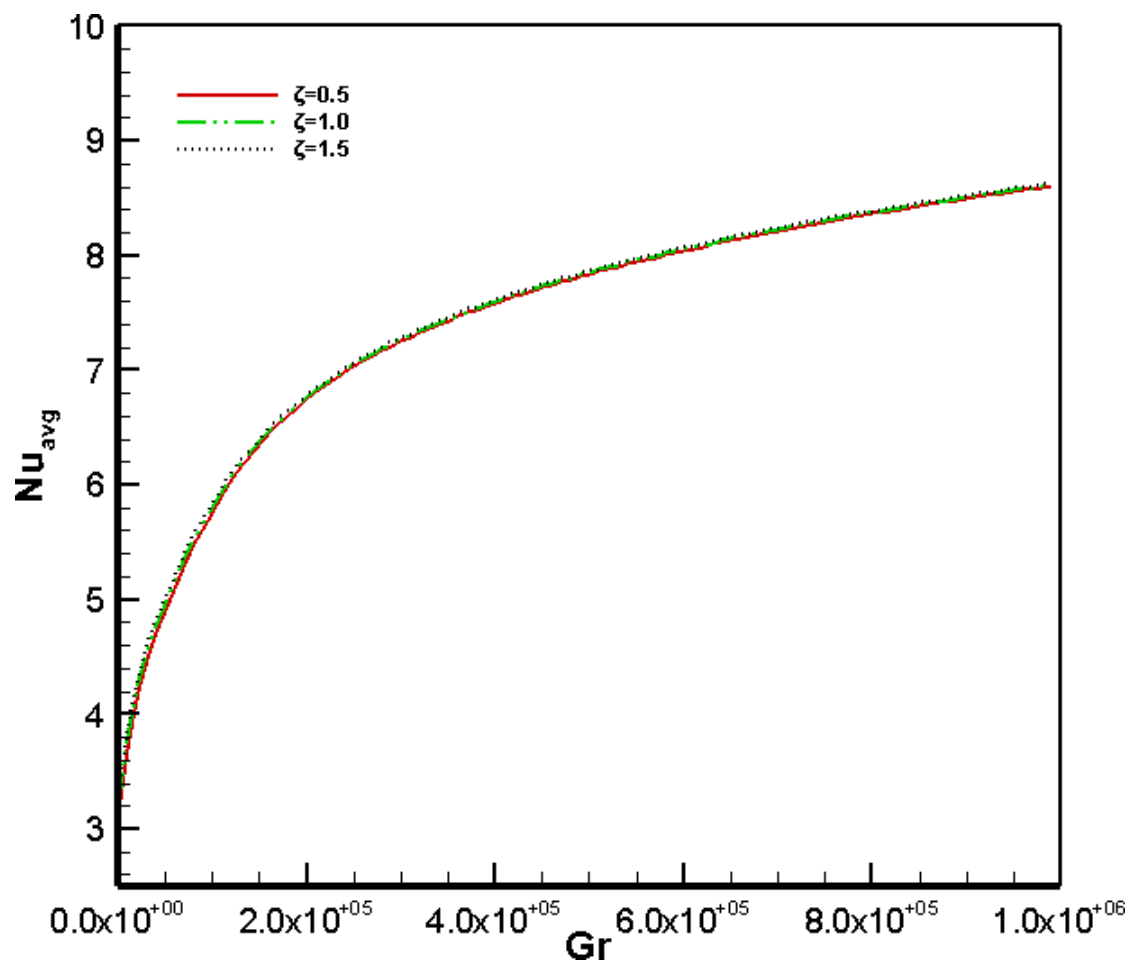

Figure 23(b). $Nu_{avg}$ vs $Gr$ at $Re = 10$ (ccw)

The local and average Nusselt number plots for various speed ratios and configurations namely: i) anti-clockwise (ccw) and clockwise (cw) rotation has been plotted in figures 22, 23, 24, 25, 26 & 27. For all cases the Gr = $10^5$for the $Nu_l$ plots and the Prandtl number of air is set to 0.71. In fig. 22(a), the local Nusselt number plots for all configurations are incident on one another, irrespective of the speed ratio. Here the local Nusselt number peaks at the left portion of the graph, falls sharply followed by a gradual decline and slightly rises at the end. This is in agreement with the isothermal line distribution plots. The surface averaged Nusselt number initially rises with Grashof number and afterwards the rate of increase falls. Similar variations of the local and average Nusselt number are obtained for the counter-clockwise case.

Similar to the previous figures, $Nu_l$ peaks at 16 in 24(a) and this occurs for the lowest speed ration whereas in the counter-clockwise case, this occurs for the highest speed ratio. This is due to increased hindrance in the convective flow field for the clockwise case with higher speed ratio. Moreover the average Nusselt number plots for Re = 50 follow a gradual rise with a decreasing rate of climb throughout the cases instead of going through a steep rise in the beginning for low Grashof numbers.The highest peak occurs for the counter-clockwise rotation of the cylinder at Reynolds number value of 100 and $\zeta$=1.5.

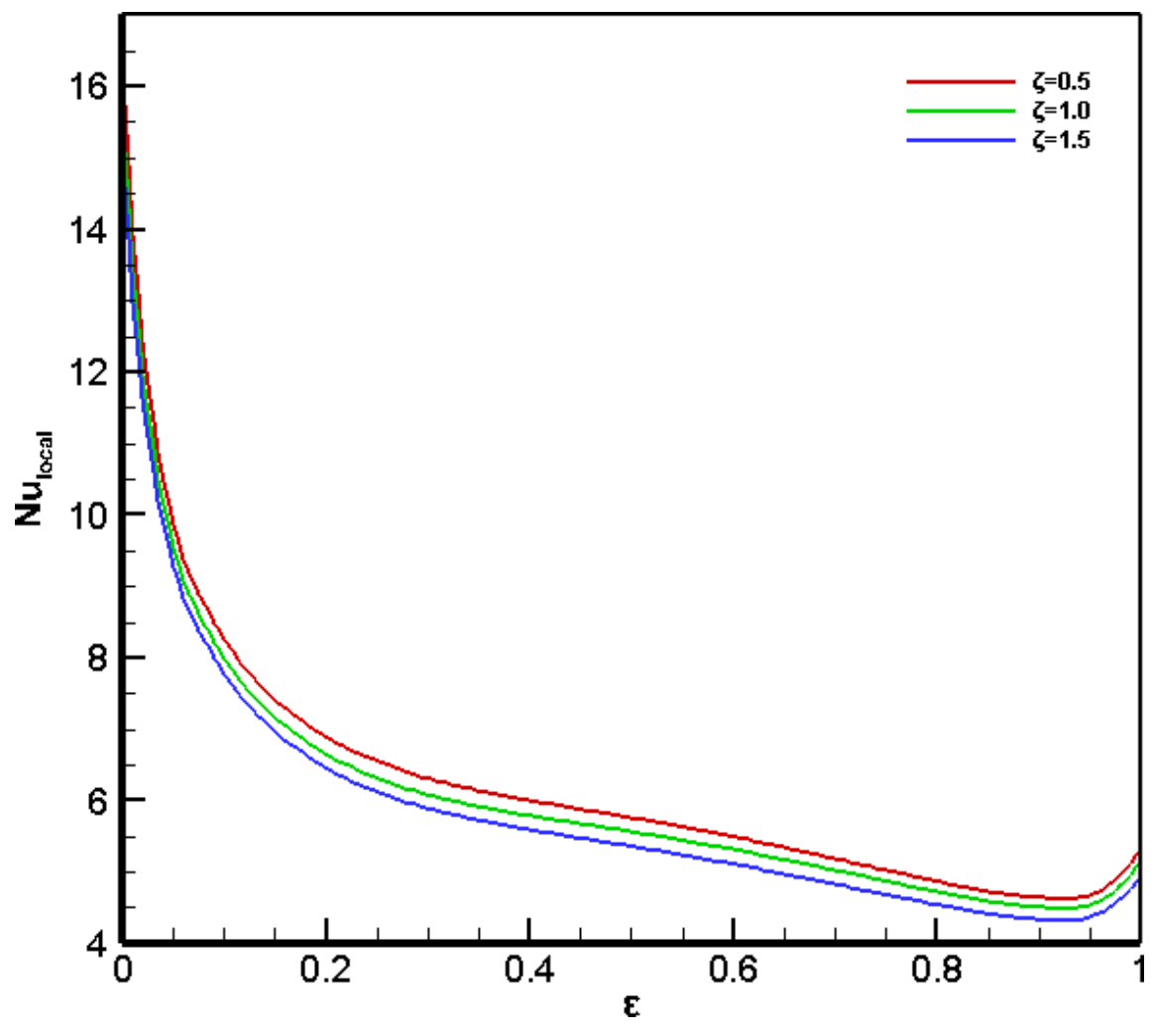

Figure 24(a). $Nu_l$ vs heater position at *Re* = 50 (cw)

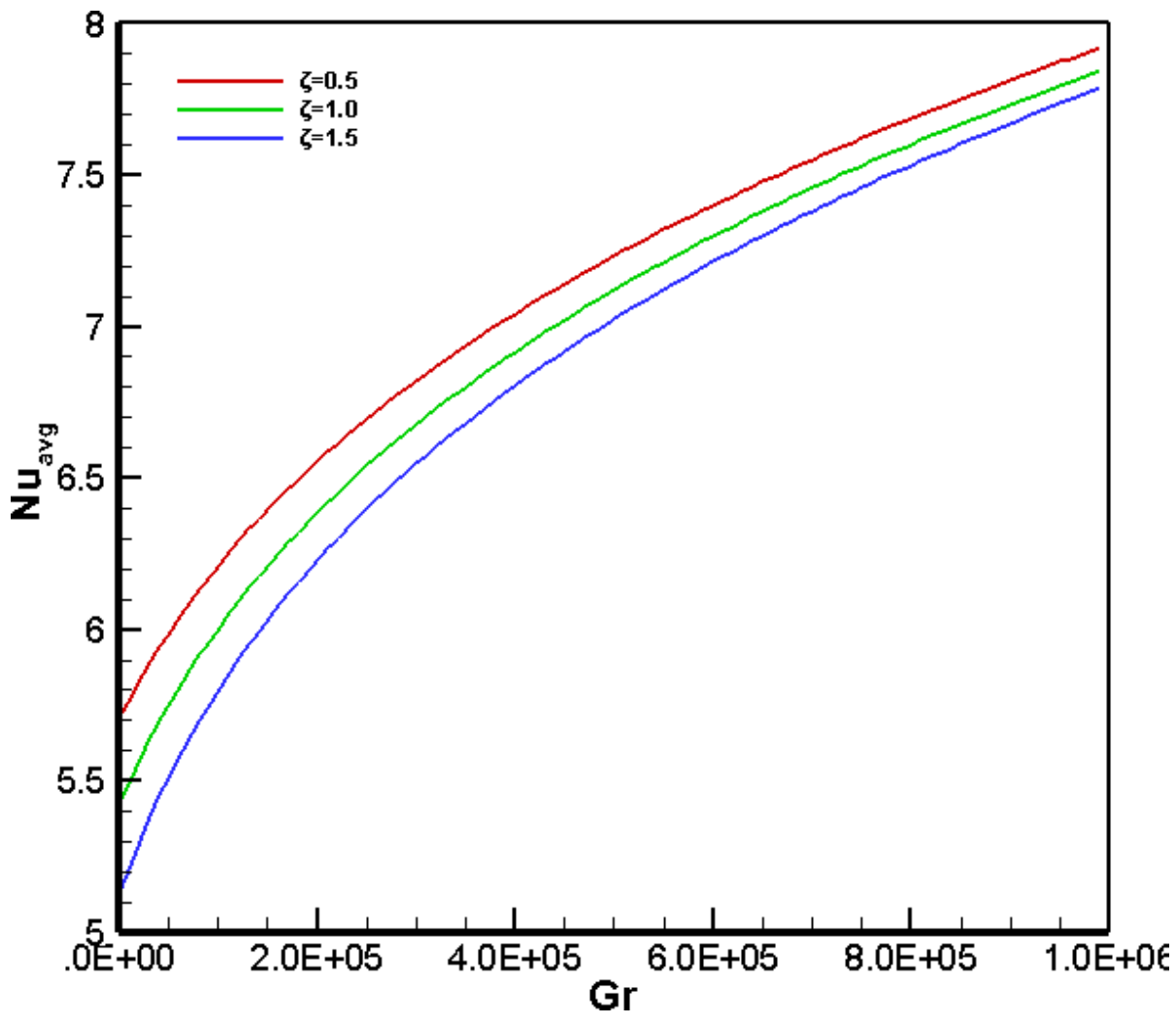

Figure 24(b). $Nu_{avg}$vs *Gr* at *Re* = 50 (cw)

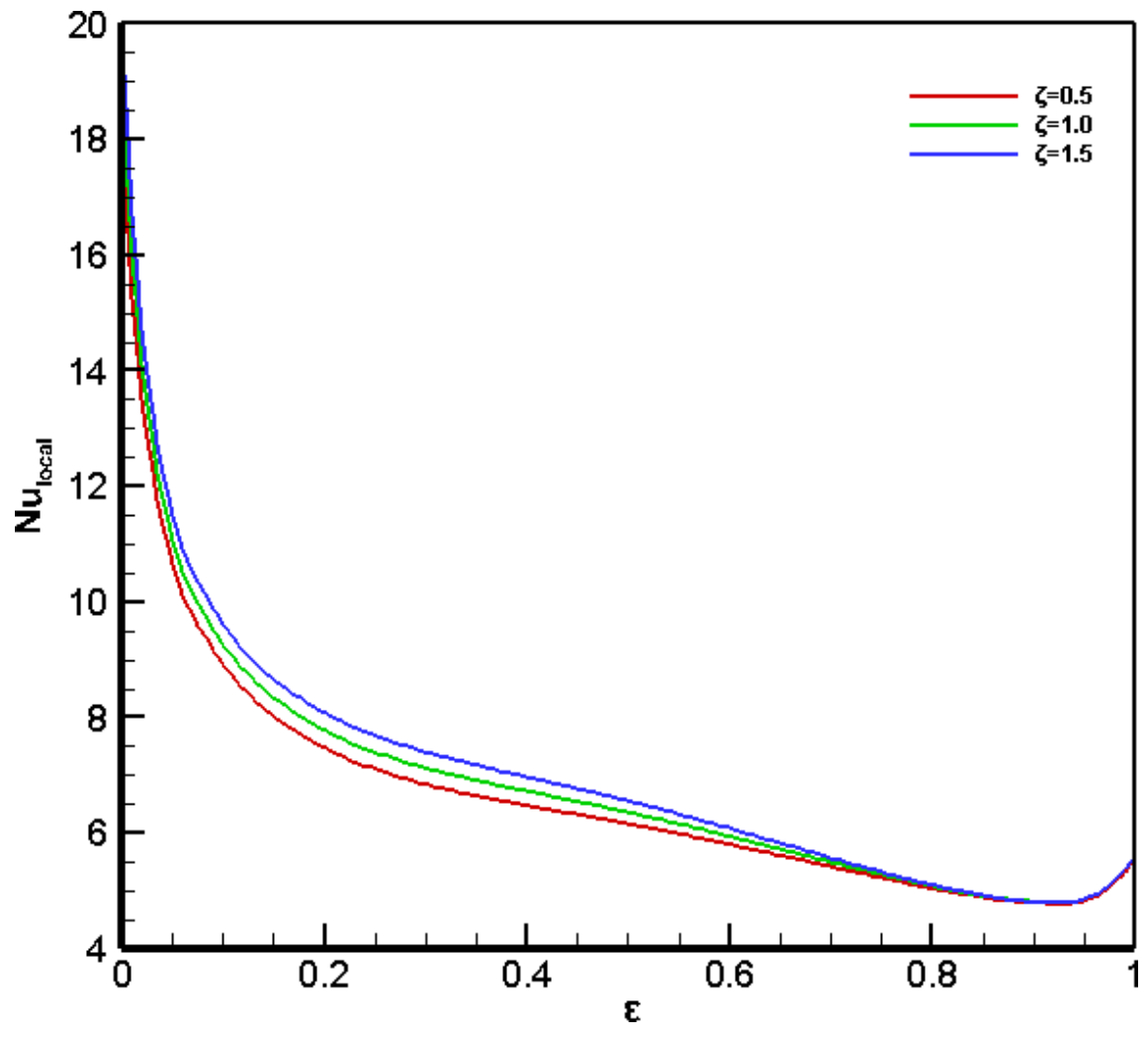

Figure 25(a). $Nu_l$ vs heater position at Re = 50 (ccw)

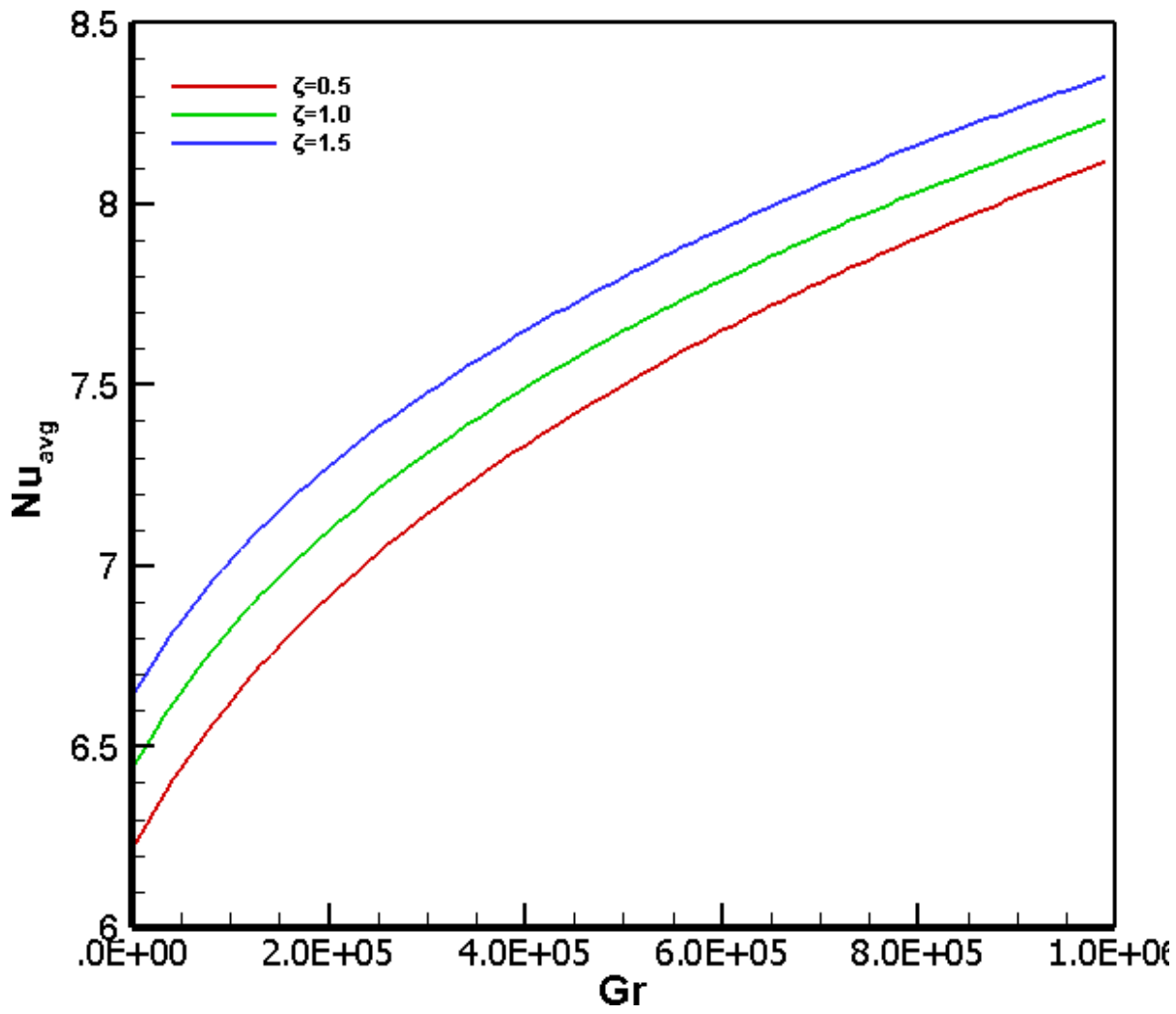

Figure 25(b). $Nu_{avg}$ *vs Gr* at *Re* = 50 (ccw)

However as seen in figure 27(a), this dips under the lower speed ratio curves towards the right end of the heater. It is interesting to see that higher speed ratios result in higher local Nusselt numbers only for the counter-clockwise rotation. This is evident in both low and high Reynolds number cases. In the average Nusselt number plots at *Re* = 50 and *Re* = 100, the average Nusselt number increases linearly with Grashof number at the start and then the rate of increase declines steadily. The strongest relationship and overall highest average Nusselt numbers are obtained for the counter-clockwise case with *Re* = 100. Similar to the local Nusselt plots, higher speed ratios enhances heat transfer for the counter-clockwise rotation case while lower speed ratios enhance

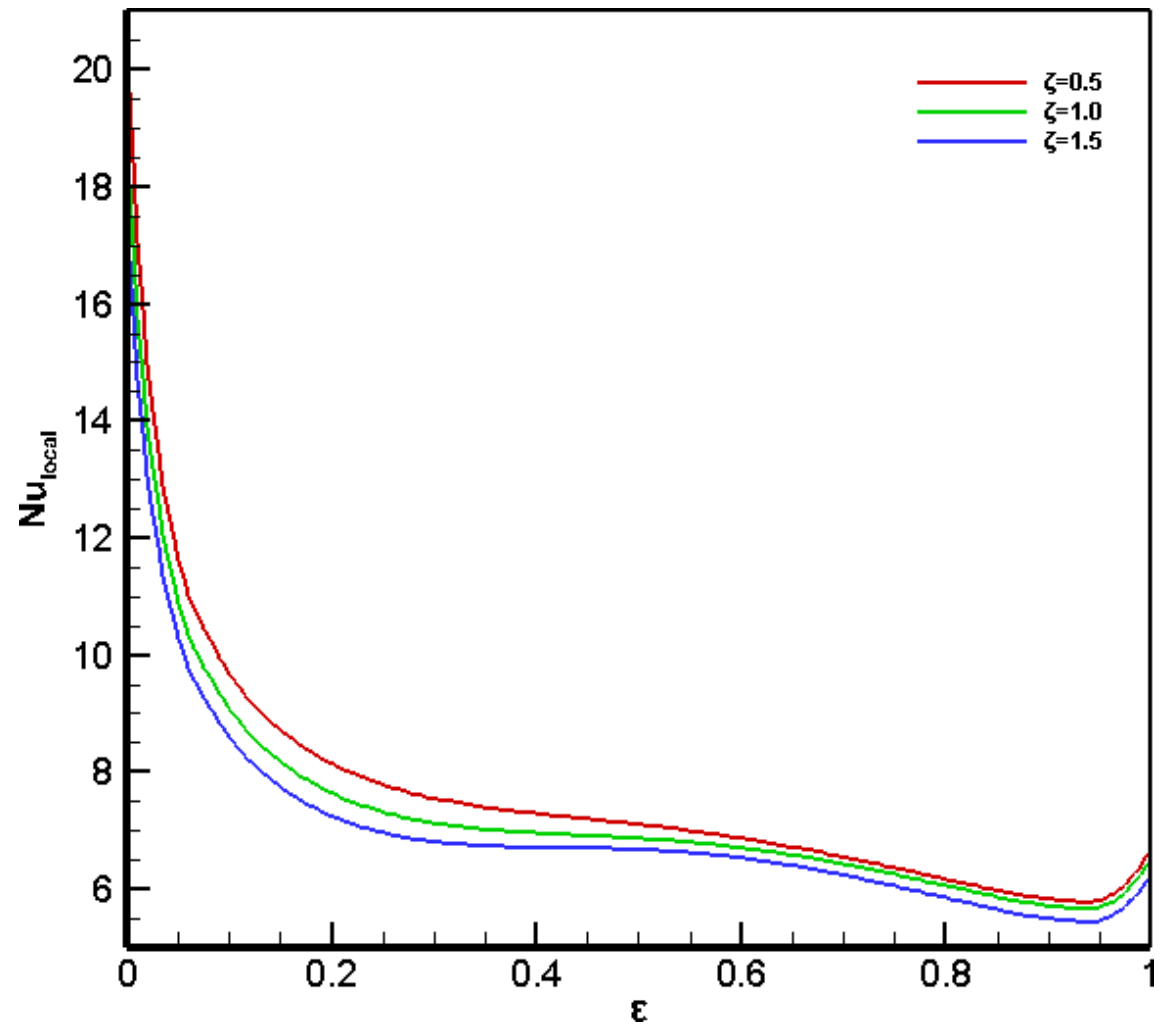

Figure 26(a). $Nu_l$ vs heater position at *Re* = 100 (cw)

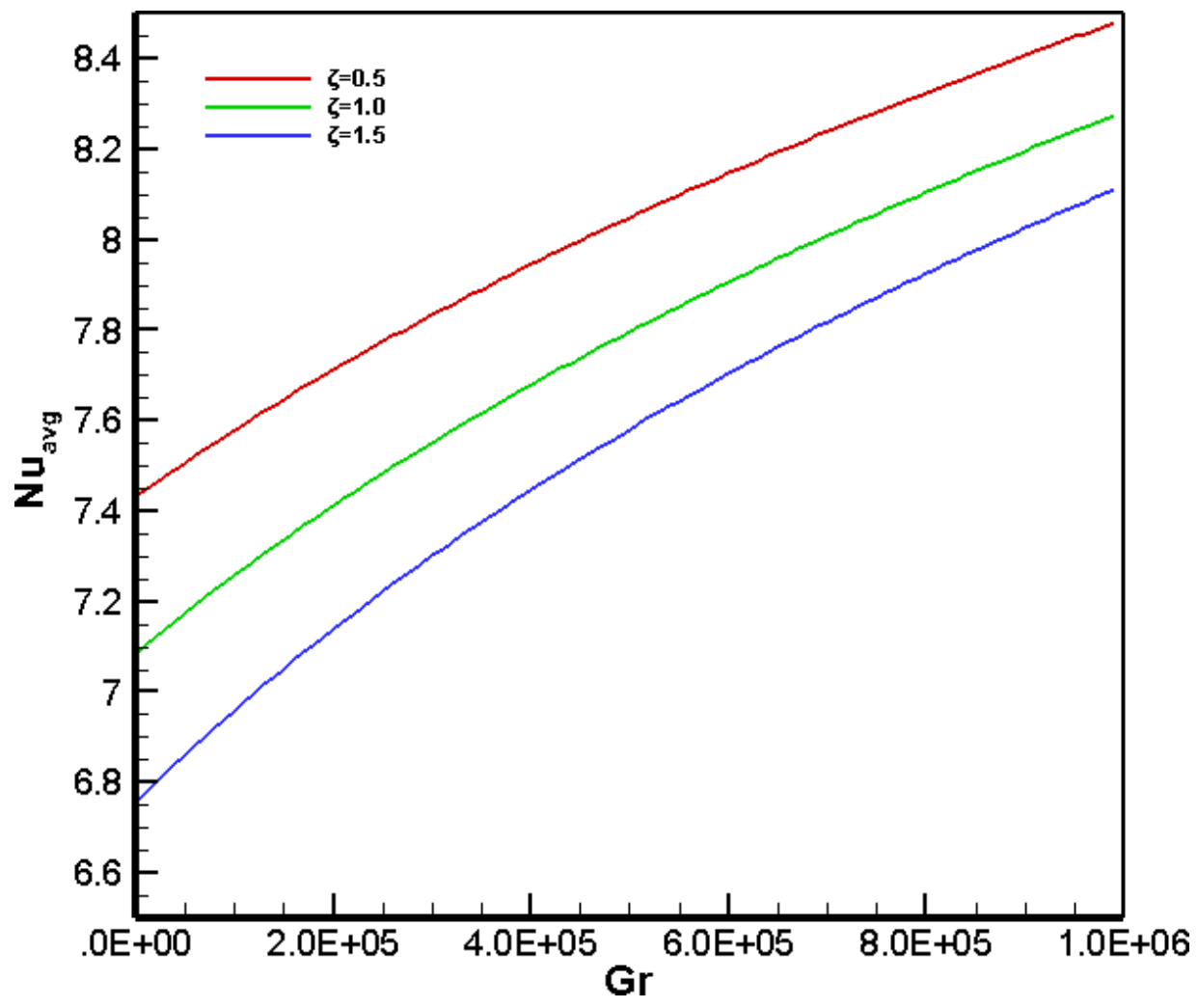

Figure 26(b). $Nu_{avg}$ vs *Gr* at *Re* = 100 (cw)

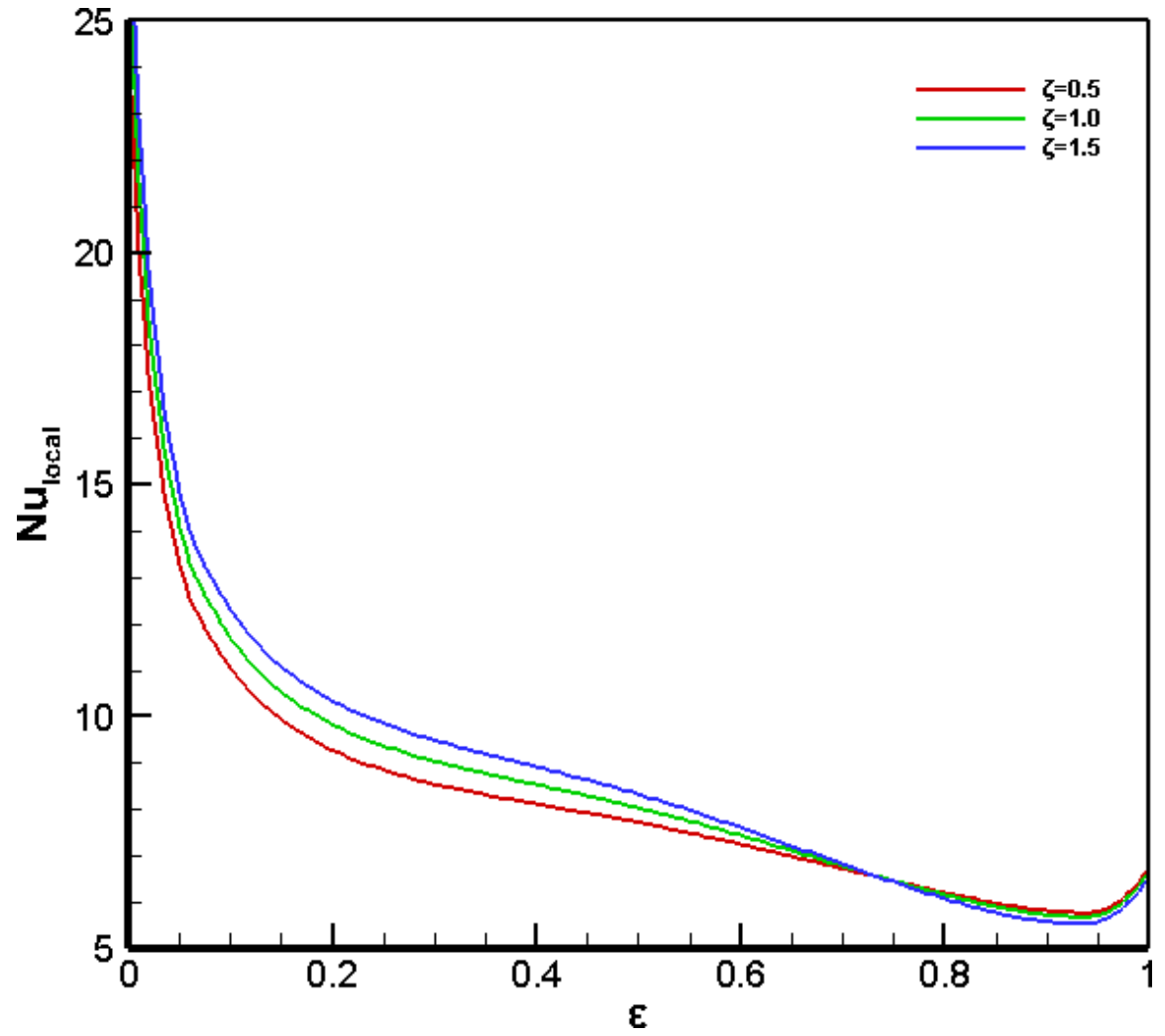

Figure 27(a). $Nu_l$ vs heater position at *Re* = 100 (ccw)

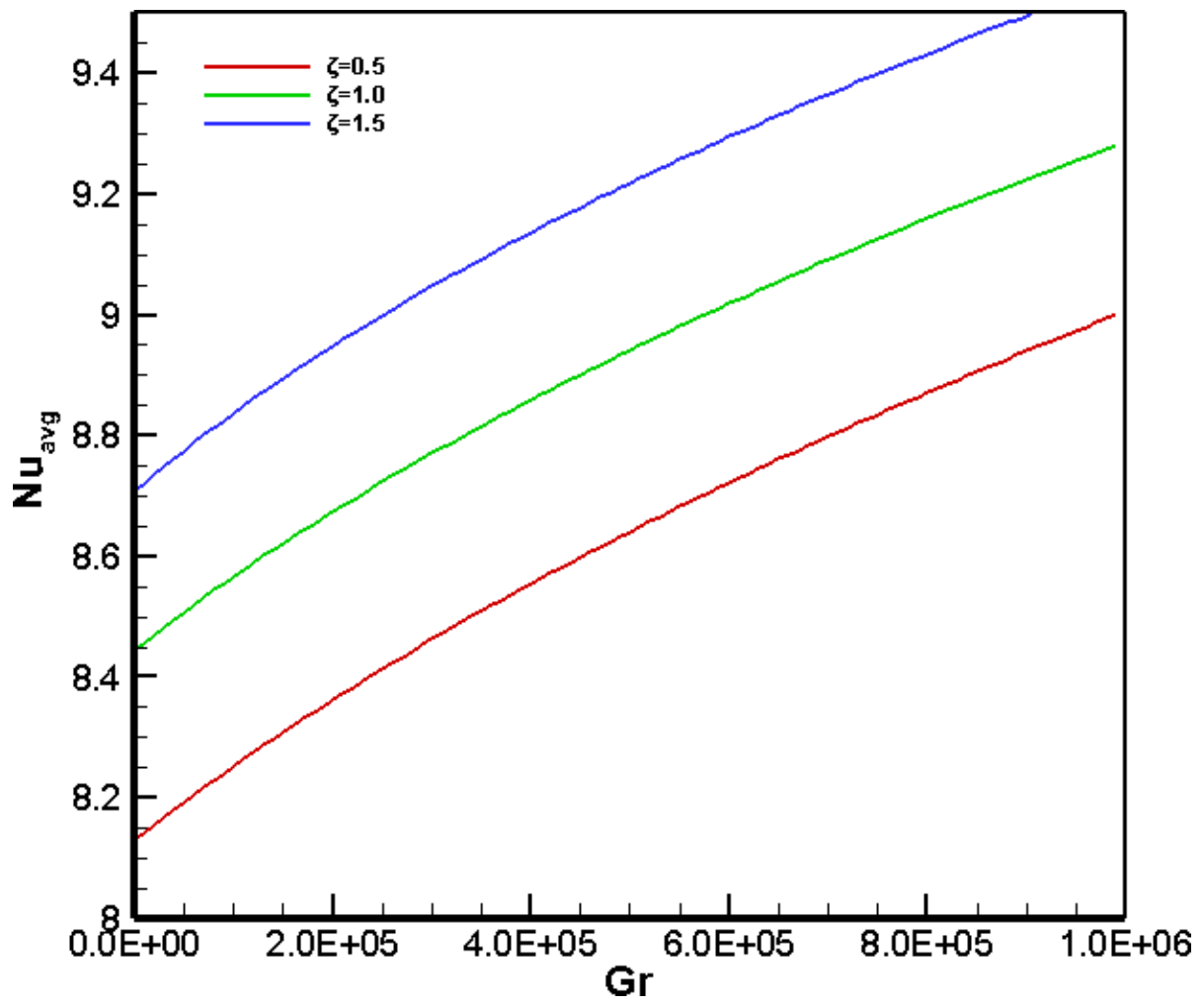

Figure 27(b). $Nu_{avg}$ vs *Gr* at *Re* = 100 (ccw)

heat transfer for the clock-wise case.

# Chapter 5

# Conclusions and Future Recommendations

## 5.1 Conclusions

The current study analyzes mixed convection heat transfer simulation in a rectangular channel having a parabolic inlet velocity profile and a variable speed heat conducting cylinder. The center of interest in this investigation is to demonstrate the effect of Grashof number, Reynolds number and cylinder speed ratio on the temperature field, flow pattern and heat transfer phenomena. These are analyzed on the basis of the distribution of isothermal lines, streamlines and local and average Nusselt numbers respectively. The following conclusions can be made from the study:

a. Nusselt number and consequently heat transfer increases for higher values of the Reynolds and Grashof numbers
b. Counterclockwise rotation arrangement of the cylinder has the highest values of local and average Nusselt numbers and thus higher rates of heat transfer.
c. Higher fluid inertia dominates the distribution of isothermal lines and leads to higher rates of heat transfer.
d. Thermal plumes are formed in various configurations with higher fluid inertia and increased Grashof number values and rate of heat transfer in these regions is high.
e. The counter-clockwise rotation case with high speed ratios and high fluid inertia demonstrated the highest rates of heat transfer.
f. Clockwise rotation case demonstrated high rates of heat transfer with low speed ratios

## 5.2Recommendations for Future Work

The prospect of this work can be further extended. Here are some recommendation for future work in this matter:

a. Instead of air, Nano fluids with different boundary conditions can be used for simulation purposes.
b. Obstacles of different geometries can be substituted in place
c. Transient phenomena can be analyzed instead of a steady state condition
d. The analysis can be carried out for turbulent flow
e. A ventilated cavity can be used instead of a rectangular channel